\documentclass[amsmath,amssymb,article,twocolumn]{revtex4}

\usepackage[version=3]{mhchem} 
\usepackage[english]{babel}
\usepackage{natbib}
\usepackage{braket}
\usepackage{gensymb}
\usepackage{graphicx}
\usepackage{courier}
\usepackage{amsmath}
\usepackage{textcomp}
\usepackage{xspace}
\usepackage{multirow}
\usepackage{soul}
\begin{document}

%%%%%%%%%%% TITLE
\title[Quantum Monte Carlo study on diradicals]{
Static and dynamical correlation in diradical molecules by Quantum Monte Carlo using the Jastrow Antisymmetrized Geminal Power ansatz
\footnote{Reprinted and adapted with permission from J. Chem. Theory Comput., DOI: 10.1021/ct401008s. Copyright 2013 American Chemical Society.}}

%%%%%%%%%%% AUTHORS
\author{Andrea Zen}
\email{zen.andrea.x@gmail.com}
\affiliation{Dipartimento di Fisica, Sapienza-Universit\`a di Roma, Piazzale Aldo Moro 5, 00185 Rome, Italy}
%{Sapienza-Universit\`a di Roma - Dipartimento di Fisica, Italy}

\author{Emanuele Coccia}
\email{emanuele.coccia82@gmail.com}
\affiliation{Dipartimento di Scienze Fisiche e Chimiche, Universit\`a degli Studi dell'Aquila, Via Vetoio 2, 67010 L'Aquila, Italy}
%{Universit\`a degli studi dell'Aquila - Dipartimento di Scienze Fisiche e Chimiche, Italy}

\author{Ye Luo}
\email{yeluo@sissa.it}
\affiliation{Scuola Internazionale Superiore di Studi Avanzati (SISSA) and Democritos National Simulation Center, Istituto Officina dei Materiali del CNR, via Bonomea 265, 34136 Trieste, Italy}
%{Scuola Internazionale Superiore di Studi Avanzati (SISSA), Italy}

\author{Sandro Sorella}
\email{sorella@sissa.it}
\affiliation{Scuola Internazionale Superiore di Studi Avanzati (SISSA) and Democritos National Simulation Center, Istituto Officina dei Materiali del CNR, via Bonomea 265, 34136 Trieste, Italy}
%{Scuola Internazionale Superiore di Studi Avanzati (SISSA), Italy}

\author{Leonardo Guidoni}
\email{leonardo.guidoni@univaq.it}
\affiliation{Dipartimento di Scienze Fisiche e Chimiche, Universit\`a degli studi dell'Aquila, Via Vetoio 2, 67010 L'Aquila, Italy}
%{Universit\`a degli studi dell'Aquila - Dipartimento di Scienze Fisiche e Chimiche, Italy}

%%%%%%%%%%%%%%%%%%%%%%%%%%%%%%%%%%%%%%%%%%%%%%%%%%%%%%%%%%%%%%%%%%%%%%
\begin{abstract}

Diradical molecules are essential species involved in many organic and inorganic chemical reactions. 
The computational study of their electronic structure is often challenging, because a reliable description of the correlation, and in particular of the static one,  requires multi-reference techniques. 
The Jastrow correlated Antisymmetrized Geminal Power (JAGP) is a compact and efficient wave function ansatz, based on the valence-bond representation, which can be used within Quantum Monte Carlo (QMC) approaches. 
The AGP part can be rewritten in terms of molecular orbitals, obtaining a multi-determinant expansion with zero-seniority number. 
In the present work we demonstrate the capability of the JAGP ansatz to correctly describe the electronic structure of two diradical prototypes: the orthogonally twisted ethylene, C$_{2}$H$_{4}$, and the methylene, CH$_{2}$, representing respectively a homosymmetric and heterosymmetric system.
In the orthogonally twisted ethylene, we find a degeneracy of $\pi$ and $\pi^{*}$ molecular orbitals,
as correctly predicted by multi-reference procedures, and our best estimates of the twisting barrier, using respectively the variational Monte Carlo (VMC) and the lattice regularized diffusion Monte Carlo (LRDMC) methods, are 71.9(1) and 70.2(2)~kcal/mol, in very good agreement  with the high-level MR-CISD+Q value, 69.2 kcal/mol. 
In the methylene we estimate 
an adiabatic triplet-singlet ($\tilde{X} ^{3}B_{1}$ - $\tilde{a} ^{1}A_{1}$) energy gap of 8.32(7) and 8.64(6)~kcal/mol, using respectively VMC and LRDMC, consistently with the experimental-derived finding for T$_{e}$, 9.363 kcal/mol. 
On the other hand, we show that the simple ansatz of a Jastrow correlated Single Determinant (JSD)  wave function is unable to provide an accurate description of the electronic structure in these diradical molecules, both at variational level 
(VMC torsional barrier of C$_2$H$_4$ of 99.3(2)~kcal/mol, triplet-singlet energy gap of CH$_2$ of 13.45(10)~kcal/mol) 
and, more remarkably,   in the fixed-nodes projection schemes 
(LRDMC torsional barrier of 97.5(2)~kcal/mol, triplet-singlet energy gap of 13.36(8)~kcal/mol) 
showing that  a  poor description of the static correlation yields an inaccurate nodal surface. 
The suitability of JAGP to correctly describe diradicals with a computational cost comparable with that of a JSD calculation, in combination with a favorable scalability of QMC algorithms with the system size, opens new perspectives in the ab initio study of large diradical systems, like the transition states in cycloaddition reactions and the thermal isomerization of biological chromophores.

\end{abstract}
%%%%%%%%%%%%%%%%%%%%%%%%%%%%%%%%%%%%%%%%%%%%%%%%%%%%%%%%%%%%%%%%%%%%%%

\maketitle

%%%%%%%%%%%%%%%%%%%%%%%%%%%%%%%%%%%%%%%%%%%%%%%%%%%%%%%%%%%%%%%%%%%%%%
\section{Introduction}\label{intro}
%%%%%%%%%%%%%%%%%%%%%%%%%%%%%%%%%%%%%%%%%%%%%%%%%%%%%%%%%%%%%%%%%%%%%%

Diradical species \cite{Salem:1972p92,Borden:1982,Bonacic:1987p170,Krylov:2006p83,Breher:2007p251} play an essential role in molecular systems, from organic reactions like cycloadditions \cite{wiest97,brad00,quad02} to chemical processes of biological interest like the  thermal isomerization of the retinal chromophore involved in the mechanism of vision \cite{pal06,gozA2012,oliv2012}.
In all cases, the reaction pathway can move through a transition state of diradical character. Archetypal examples are the dimerization of butadiene \cite{brad00,quad02} and the reaction between butadiene and ethylene \cite{wiest97}, in which diradical transition states are supposed to be energetically competitive with aromatic ones.  As example, the ground state potential energy surface of the retinal chromophore,\cite{oliv2012} and of its reduced model\cite{gozA2012,xu13} on which the thermal isomerization occurs, is characterized by two different branches, one with a charge-transfer character and the other of diradical character. To quantitatively define the energy landscape of such paths, very high level calculations are necessary, including an adequate treatment of both static and dynamic electronic correlation. On the other hand, determining the relative stability of singlet and triplet states can represent a challenging task for theory even for small molecules like the tetramethyleneethane\cite{Pozun:2013hy}.  Spin multiplicity is also a fundamental ingredient in the design of magnetic materials involving diradical species, like Zn-expanded oligoacenes.  \cite{abe2013}\\
According to Salem and Rowland's definition \cite{Salem:1972p92}, a diradical species is a molecule with two unpaired electrons occupying two (near-)degenerate molecular orbitals. 
Molecules with a broken double bond represent a model for diradicals. One of the simplest cases is the torsion around the double carbon-carbon bond of ethylene,  C$_2$H$_4$. \cite{doug55,rabi59,Gemein:1996jr,Sension:1987gc,wal89,krylov98,Gwaltney:2000fj,krylov00,krylov01,Shao:2003p4807,Barbatti:2004p11614,Krylov:2006p83,casanova08,Lopez:2011p1673,maha11}  Drifting away from the planar structure to the orthogonally twisted conformation,  the bonding $\pi$ orbital and the anti-bonding $\pi^{*}$ orbital of the singlet ground state become degenerate and a single-configuration $\pi^{2}$ wave function is not appropriate anymore to describe the electronic state: the orthogonally twisted ethylene represents the prototype of the homosymmetric diradical. \cite{Salem:1972p92} 
Standard density functional theory (DFT) and restricted  Hartree-Fock (RHF) approaches indeed fail even in the qualitative description of the shape of the torsional barrier (cusp at 90\textdegree~instead of a smooth profile). \cite{krylov98,Shao:2003p4807,Lopez:2011p1673} 
In order to describe the twisted ethylene at 90\textdegree~, at least two configurations $\pi^{2}$ and $\pi^{*2}$ have to be included. 

Following the classification reported in Ref. \citenum{Salem:1972p92},   the methylene molecule, CH$_{2}$,\cite{dav80,Reynolds:1985,handy86,scha86,jensen88,alijah90,piecuch94,woon95,Yamaguchi:1996p7911,Sherrill:1998p1040,Slipchenko:2002p4694,li08,shen08,Zimmerman:2009hh,gour10,and2010,shen12} is considered a heterosymmetric diradical, since the two orbitals occupied by the unpaired electrons have different symmetries (1b$_{1}$ and 3a$_{1}$) \cite{Salem:1972p92}.   
CH$_2$ was extensively studied to assess  the reliability of quantitative ab initio calculations for the triplet geometry and the singlet-triplet ($\tilde{X} ^{3}B_{1}$ - $\tilde{a} ^{1}A_{1}$) energy gap\cite{scha86}.

Both the static and the dynamical correlation play an essential role in the description of the electronic structure of diradicals, from benchmark systems like ethylene, methylene or tetramethyleneethane \cite{Pozun:2013hy} to complex transition states or reaction intermediates in biological processes. On one side, if single-reference methods are inadequate in treating the quasi-degeneracy of singly occupied orbitals in diradicals, on the other side multiconfigurational approaches suffer from some limitations due to the computational cost of the multideterminantal expansion of the wave function. The aim of the present paper is to demonstrate that  Quantum Monte Carlo (QMC) is a valid alternative to the traditional quantum chemistry methods for the correct description of diradical molecules.

QMC methods\cite{Foulkes:2001p19717,Austin:2012kt} 
have been successfully applied to various fields of physics and chemistry, like the study of molecular properties \cite{Caffarel:1993td,Schautz:2004p25285,Sorella:2007p12646,Sterpone:2008p12640,Zimmerman:2009hh,Barborini:2012iy,Zen:2012br,Coccia:2012fi},  materials \cite{Spanu:2009p12613, Maezono:2010ic,lib2011a,lib2011b,Kolorenc:2011hv,Mazzola:2012ch}, reaction pathways \cite{Barborini:2012it,saccani13} and  biomolecules \cite{Valsson:2010p25419,Filippi:2012hg,Coccia:2012kz,Coccia:2012ex}.  
They are characterized by a good scalability with respect to the system size ($N^d$, with $3<d<4$ and $N$ the number of electrons)\cite{Foulkes:2001p19717,Coccia:2012kz}, comparable with that of DFT, and by using algorithms that can be efficiently parallelized (making them extremely suitable for the PetaScale architectures). These ingredients justify the growing number of applications of QMC in problems of quantum chemistry or molecular physics.  
% VMC
The variational Monte Carlo (VMC) method\cite{bres98}  
exploits the combined use of Monte Carlo integration and variational principle to optimize the many-body, ground-state trial wave function. 
% FN-QMC
Further improvements are given by using the fixed node projection Monte Carlo methods, such as the diffusion Monte Carlo (DMC)\cite{Reynolds:1982en,bk:hammond} and the lattice regularized diffusion Monte Carlo\cite{Casula:2010p14082} (LRDMC).

% JAGP
The wave function ansatz is the ingredient that mostly affects the accuracy of QMC approaches, both in the variational and  in the fixed node projection schemes (although in the latter the effect is alleviated).
We propose here the Jastrow Antisymmetrised Geminal Power (JAGP)\cite{Casula:2003p12694,Casula:2004p12689, Neuscamman:2012hm, Neuscamman:2013a, Neuscamman:2013b, Stella:2011} trial wave function as a promising ansatz to study diradical systems. 
 The JAGP wave function, which implements Pauling's resonating valence bond idea\cite{bk:pauling},   
 has been seen to be very efficient in the study of problems of chemical interest, \cite{Barborini:2012iy,Zen:2012br,Coccia:2012fi, Barborini:2012it,  Coccia:2012kz,Coccia:2012ex} with accuracy comparable to that of high-level quantum chemistry methods. 
Its compactness, combined with the use of efficient algorithms for the optimization of all parameters, including linear coefficients and exponents of the atomic basis sets,\cite{Sorella:2005p14143, Sorella:2007p12646,Umrigar:2007p12662} leads to a fast convergence of the variational results for electronic and geometrical properties with the size of the basis sets\cite{Petruzielo:2011p24345,Barborini:2012iy,Zen:2013is}, with a computational cost comparable to that of a simplest Jastrow Single Determinant (JSD). With a large Jastrow factor JAGP is size consistent\cite{Sorella:2007p12646,Marchi:2009p12614,Neuscamman:2012hm}, without any spin contamination, for partitioning the system in fragments of spin zero and of spin 1/2. JAGP has already proven to give a good description of the static and dynamic correlation in several cases\cite{Stella:2011et,Neuscamman:2013a,Neuscamman:2013b},
and have inspired new ansatzes such as the 
linearized Jastrow-style fluctuations on spin-projected Hartree-Fock\cite{Henderson:2013gf}.

% FINE ED OUTLINE 
In the present work, the multiconfigurational nature of JAGP is explicitly described, with specific attention to the determination of formal analogies with standard molecular orbital theories for two benchmark systems: the ethylene (C$_{2}$H$_{4}$) and the methylene (CH$_{2}$), representing examples respectively of homosymmetric and heterosymmetric diradicals. 
We show how accurate and reliable QMC calculations based on the JAGP ansatz are in the estimation of the torsional barrier of C$_{2}$H$_{4}$ and of the triplet-singlet gap in CH$_{2}$.
The paper is organized as follows: 
in Section \ref{qmc} we report the basic concepts of QMC, focusing the attention on the wave function ansatz and providing a detailed description of the JAGP wave function, in comparison with other multideterminat approaches and Salem and Rowland's two-electron model for the description of diradical molecules.
The computational details for QMC and CASSCF calculations reported in this work are given in Section \ref{cdet}.
The results on the torsional barrier of the orthogonally twisted ethylene and on the singlet-triplet gap of the methylene are reported and discussed in Section \ref{res}. 
Final remarks are reported in the Conclusions,
highlighting how this work constitutes a fundamental step for using QMC  in order to have an accurate and computationally feasible description of several chemical processes of biological interest like  the thermal isomerization of the retinal chromophore in the mechanism of vision.

%%%%%%%%%%%%%%%%%%%%%%%%%%%%%%%%%%%%%%%%%%%%%%%%%%%%%%%%%%%%%%%%%%%%%
\section{Quantum Monte Carlo and wave function ansatzes}
%%%%%%%%%%%%%%%%%%%%%%%%%%%%%%%%%%%%%%%%%%%%%%%%%%%%%%%%%%%%%%%%%%%%%  
\label{qmc}
%%%%%%%%%%%%%%%%%%%%%%%%%%%%%%%%%%%%%%%%%%%%%%%%%%%%%%%%%%%%%%%%%%%%%%

% QMC w.f.
The accuracy of  QMC approaches, both in the simplest VMC scheme and in the fixed-node projection schemes, are strictly related to the wave function ansatz.
Typically, the electronic wave function $\Psi_{T}$ in QMC \cite{Foulkes:2001p19717,Austin:2012kt,Zen:2013is} is  defined by the  product 
\begin{equation}
\Psi_{T}(\bar{\mathbf{x}};\bar{\mathbf{R}}) = {\cal D}(\bar{ \mathbf{x}};\bar{ \mathbf{R}}) {\cal J}(\bar{ \mathbf{x}};\bar{ \mathbf{R}}) ,
\end{equation}
where $\cal D$ is the antisymmetric function taking into account the fermionic nature of electrons and $\cal J$ is the Jastrow factor depending on inter-particle (electrons and nuclei) distances; $\bar{ \mathbf{x}}$ and $\bar{ \mathbf{R}}$ represent the collective electronic ($\bar{ \mathbf{x}}$ refers to space $\bar{ \mathbf{r}}$ and spin $\bar{ \sigma}$) and nuclear coordinates, respectively. 
The Jastrow factor is a  symmetric positive function of the electronic positions; therefore it does not change the nodal surface (determined by the antisymmetric term $\cal D$), but it describes the dynamical correlation among electrons and satisfies the electron-electron and electron-nucleus cusp conditions \cite{Foulkes:2001p19717,drum04,Zen:2013is}. 

% VMC
In VMC, the parameters that define $\Psi_{T}$ are optimized, within the functional freedom of the ansatz, in order to minimize the electronic energy. The variational principle ensures that wave function ansatzes which give a lower energy provide a better description of the ground state electronic structure.
Some of the limitations of the VMC approach, ultimately ascribable to the wave function ansatz,
 can be alleviated by adopting the fixed-node (FN) projection Monte Carlo techniques, which provide the lowest possible energy, with the constraint that the wave function $\Phi_{FN}$ has the same nodal surface of an appropriately chosen guiding function $\Psi_G$ ({ fixed node approximation})\cite{Reynolds:1982en,Foulkes:2001p19717}, that typically is the variationally optimized function $\Psi_T$.
% LRDMC
In this work  we have used  the lattice regularized diffusion Monte Carlo\cite{Casula:2005p14138,Casula:2010p14082}:
it turns out to be an efficient scheme even for systems with a large number of electrons\cite{Casula:2010p14082}
and it preserves the variational principle even when used in combination with nonlocal pseudopotentials\cite{Casula:2010p14082}.
Moreover, the extrapolation for mesh size $a\to 0$ is generally easier than the extrapolation of  time step $\tau \to 0$ in  ordinary DMC\cite{Casula:2010p14082}.

% JAGP
The ansatzes considered in this work are the Jastrow correlated Antisymmetrized Geminal Power (JAGP, product $\Psi_{JAGP} =  \Psi_{AGP} \cdot J$ of an Antisymmetric Geminal Power and a Jastrow factor $J$), the JAGP with a fixed number $n$ of molecular orbitals (JAGPn, product $\Psi_{JAGPn} =  \Psi_{AGPn} \cdot J$) and the Jastrow correlated Single Slater Determinant (JSD, product $\Psi_{JSD} =  \Psi_{SD} \cdot J$).

% riassuntino 
In Paragraph~\ref{sec:Jas} we review the main features of the Jastrow factor, the same in Paragraph~\ref{sec:AGP}  is done for the AGP and AGPn.
We show in Paragraph~\ref{sec:multidetAGP} that the JAGP ansatz is intrinsically multiconfigurational, yielding an improvement of the JSD ansatz, mainly in terms of static correlation.
Finally, in Paragraph~\ref{sec:JAGP_diradicals} we show that the JAGP function includes the expected leading ingredients for a reliable description of the diradicals.

%%%%%%%%%%%%%%%%%%%%%%%%%%%%%%%%%%%%%%%%%%%%%%%%%%%%%%%%%%%%%%%%%%%%%
\subsection{The Jastrow factor} \label{sec:Jas}

The implementation of the Jastrow factor
adopted in this paper is described extensively in Ref.~\citenum{Zen:2013is}.
The Jastrow factor $J=e^U$ used in our calculations consists of several terms accounting for the 2-body, 3-body and 4-body interaction between the electrons and the nuclei.
The exponent $U$ of the Jastrow factor can therefore  be conveniently written as the sum 
$$ U = U_{en} + U_{ee} + U_{een} + U_{eenn} $$
of the electron-nucleus function $U_{en}$, 
the electron-electron function $U_{ee}$,
the electron-electron-nucleus function $U_{een}$ and 
the electron-electron-nucleus-nucleus function $U_{eenn}$.
The leading contribution is given by 
$U_{ee}$, which yields a homogeneous \textit{two-electron} interaction term.
It depends only on the distance between pairs of electrons and it improves the electron-electron correlation, and it is used to satisfy the electron-electron cusp condition. 
The  \textit{one-electron} interaction term $U_{en}$   improves the electron-nucleus correlation and satisfies the nuclear cusp condition. 
The  $U_{een}$ and $U_{eenn}$ functions describe an inhomogeneous {\em two-electron} interaction, which further correct the correlation introduced by the homogeneous term $U_{ee}$ (and within the JAGP ansatz it  reduces the unphysical charge fluctuations included in the AGP function\cite{Sorella:2007p12646,Neuscamman:2012hm}).

Actually, the electron-electron coalescence problem  and the corresponding form of the $U_{ee}$ term need a further explanation.
Pairs of like-spin electrons have to satisfy a cusp condition that is different from that of unlike-spin electrons\cite{Foulkes:2001p19717}. 
As shown in Refs.~\cite{Filippi:1996eb,Casula:2003p12694}, if two different electron terms are used to satisfy respectively the like-spin and the unlike-spin cusp conditions, this would translate in a spin contaminated wave function.
Thus, two approaches are possible: 
({\em i}) to satisfy only the cusp condition for unlike-spin, because the probability for like-spin electrons to be close is very small, according to the Pauli principle;
({\em ii}) to satisfy both the cusp conditions, by using $U_\textrm{like}$ and $U_\textrm{unlike}$ respectively for like and unlike spin electrons.
In this paper, unless explicitly stated, we have adopted the solution ({\em i}), using the following functional form for the homogeneous two electron interaction term:
% 2bJ
\begin{equation}\label{equ:2BJas}
U_{ee}\left(\bar{\textbf{r}}\right) = 
  \sum_{i<j}^N { 1-\exp({- b r_{ij}}) \over 2 b }
\end{equation}
where $r_{ij}=\|\textbf{r}_{i}-\textbf{r}_{j}\|$ is the distance between electrons $i$ and $j$, 
the summation runs over all the $N$ electrons in the system irrespectively from their spin states, and 
$b$ is a variational parameter.
In this case we will use $J$ to indicate the corresponding ansatz.  
% spin contamination
However, in some cases we have adopted the solution ({\em ii}), by using:
\begin{eqnarray}\label{equ:2BJas-sc}
U_{ee}\left(\bar{\textbf{r}}\right) =& 
  \sum_{i<j}^{N} \left[
    (1-\delta_{\sigma_i,\sigma_j}) { 1-\exp({- b r_{ij}}) \over 2 b } \right. \nonumber \\
    &  \left. +
    \delta_{\sigma_i,\sigma_j} { 1-\exp({- b r_{ij}}) \over 4 b } 
  \right]
\end{eqnarray}
where $\sigma_i$ indicates the spin state of electron $i$, and  $\delta_{\sigma_i,\sigma_j}$ is the Kronecker's delta function.
In these cases, the ansatz will be indicated with $J^*$. 
The spin contamination induced by this ansatz  can be evaluated by calculating the total spin of the molecule $\left< S^2 \right>$, see Appendix~\ref{app:s2}. 

%%%%%%%%%%%%%%%%%%%%%%%%%%%%%%%%%%%%%%%%%%%%%%%%%%%%%%%%%%%%%%%%%%%%%
\subsection{The AGP and AGPn functions}\label{sec:AGP}

A spin-unpolarized  molecular system of $N=2N_p$ electrons and $M$ nuclei describes a singlet state that, within the AGP ansatz, is written as:
\begin{equation}\label{equ:AGP}
\Psi_{AGP}\left(\bar{\textbf{x}}\right) = 
  \hat{\cal A} \left[ 
     \prod_{i}^{N_p} G \left( \textbf{x}_{i};\textbf{x}_{N_p+i} \right) 
  \right] ,
\end{equation} 
where $\hat {\cal A}$ is the antisymmetrization operator, and the geminal pairing function $G$ is a product of a singlet function and a symmetric spatial wave function $\cal G$:
\begin{equation}\label{equ:G}
G(\textbf{x}_{i} ;\textbf{x}_{j}) = 
   {\cal G}\left( \textbf{r}_{i},\textbf{r}_{j} \right) 
   \frac{ \alpha(i) \beta(j) - \beta(i) \alpha(j) }{\sqrt{2}}.
\end{equation}
The spatial function $\cal G$ is a linear combination of products of atomic orbitals $\phi_\mu$: 
\begin{equation}\label{equ:calG}
{\cal G}\left( \textbf{r}_{i},\textbf{r}_{j} \right) = 
  \sum_{\mu}^{L} \sum_{\nu}^{L} g_{\mu \nu} \phi_{\mu}\left(\textbf{r}_{i}\right)\phi_{\nu}\left(\textbf{r}_{j} \right)
\end{equation} 
where the indexes $\mu$ and $\nu$ run over all the basis in all the atoms in the system, for a total of $L$ atomic orbitals (note that $L$ is determined by the overall basis set size).
% N. parameters
The coefficients $g_{\mu\nu}$ have to be optimized in order to minimize the variational energy of the system (together with the other parameters in the wave function).
They define a symmetric (because $\cal G$ is symmetric) $L \times L$ matrix $\bf G$, for a total of $L(L+1)/2$ variational parameters (unless the system is characterized by some  symmetry property reducing the total number of parameters, see Ref.~\citenum{Casula:2004p12689}).

%%% MOs per fare AGPn
%
The pairing function $\cal G$ in Eq.~(\ref{equ:calG}) is written in terms of the (localized) atomic orbitals $\phi_\mu$, offering an interesting correspondence between the AGP ansatz and the Resonating Valence Bond framework \cite{Casula:2005p14146,Sorella_book2013}. 
%
% MOs
An equivalent way to write the pairing function $\cal G$ is obtained by using the molecular orbitals (MOs) $\psi_k$. % 
The expansion of the pairing function in terms of MOs is obtained by performing a generalized (the atomic orbitals $\phi_\mu$ are not necessarily orthonormal, so the overlap matrix $S_{\mu\nu} =\left<\phi_{\mu}|\phi_\nu \right> \ne \delta_{\mu\nu}$) diagonalization of  the $\bf G$ matrix  appearing in Eq.~(\ref{equ:calG}):
\begin{eqnarray} \label{gendiag}
 & {\bf G S P}  = {\bf P} \Lambda \\
 & \textrm{where   } {\Lambda} = \text{diag}(\lambda_1,\ldots,\lambda_L) \nonumber \\
 & \textrm{and   } |\lambda_1| \ge |\lambda_2| \ge \ldots \ge |\lambda_L |\ge 0 . \label{equ:lambdas}
\end{eqnarray} 
In Eq.~(\ref{gendiag}) each column of the matrix $\bf P$ represents a generalized eigenvector of $\bf G$ and $\bf S$ is the overlap matrix.
Thus, from $ {\bf P}^T {\bf S P} = {\bf 1}$, by right multiplying both sides of Eq. (\ref{gendiag})  for the 
matrix ${\bf P}^T= ({\bf SP})^{-1}$ we obtain   
${\bf G} = {\bf P}  { \Lambda}  {\bf P}^T$.
Then, by substituting it in Eq.~(\ref{equ:calG}), we finally obtain that 
the pairing function is:
\begin{equation}
{\cal G} \left( \textbf{r}_{i},\textbf{r}_{j} \right) =
  \sum_{k=1}^{L}  
    \lambda_{k} 
    \psi_{k}\left(\textbf{r}_{i}\right)
    \psi_{k}\left(\textbf{r}_{j} \right),
\label{equ:G_MOs}
\end{equation} 
where the orthonormal single particle functions
$ \psi_k(\textbf{r}) = \sum_{\mu=1}^{L} P_{\mu k} \phi_{\mu}(\textbf{r}) $
are the molecular orbitals.
Considering this expansion   of the pairing function ${\cal G}$, 
it may be reasonably expected that the leading terms are provided by the MOs associated to the largest (in absolute value) eigenvalues $\lambda_k$.
Thus, by considering a truncated pairing function 
\begin{equation}\label{equ:gn}
{\cal G}_n\left( \textbf{r}_{i},\textbf{r}_{j} \right) =
  \sum_{k=1}^{n}  
    \lambda_{k} 
    \psi_{k}\left(\textbf{r}_{i}\right)
    \psi_{k}\left(\textbf{r}_{j} \right)
\end{equation}
with $N_p\le n\ll L$, 
the quality of the parametrical wave function is not significantly affected if $n$ is large enough, but a substantial reduction of the number of variational parameters is obtained.
If this truncated pairing function ${\cal G}_n$ is used in place of ${\cal G}$ in the AGP, we have what in this paper is named the AGPn function.
A particular case of AGPn is for $n=N/2$, that reduces to a single Slater Determinant function, as it will be proved in the next section.

% generalized AGP
In order to describe a polarized system, with total spin $S$, 
a generalized AGP (GAGP) should be used.\cite{Casula:2003p12694}
The system is constituted by $N_p$ paired electrons, and $N_u\equiv 2 S$ unpaired electrons with same spin, that without loss of generality can be considered spin-up.
Thus the system has $N_{\uparrow} = (N_p + N_u)$ spin-up electrons, and 
$N_{\downarrow} = N_p$ spin-down electrons, for a total of $(2 N_p +N_u)$ electrons.
The GAGP wave function is written as follows:
\begin{equation}
\Psi_{GAGP}\left(\bar{\textbf{x}}\right) = 
   \hat{\cal A} \left\{ \left[ 
      \prod_{i}^{N_{p}} G\left( \textbf{x}_{i};\textbf{x}_{N_p+i} \right) 
      \right] \left[ 
      \prod_{j}^{N_u} {\chi}_j \left( \textbf{x}_{2 N_p + j} \right) 
      \right] \right\} ,
\label{eq:gagp1}
\end{equation}
where single-electron functions 
${\chi}_j$, with $j=1,\ldots,N_u$, have been introduced.
The generic function $\chi_j$ is written as:
\begin{equation}
{\chi}_j ( \textbf{x}_i) = 
   \left[ \sum_{\mu}^{L} f_{j,\mu} \phi_{\mu} (\textbf{x}_i) \right] \alpha(i) ,
\label{eq:gagp2}
\end{equation}
where the $f_{j,\mu}$ coefficients are variational parameters.
Each unpaired electron requires the addition of $L$ variational parameters, for a total of 
$L(L+1)/2 + 2 S L$ determinantal parameters for the GAGP function of a system with total spin $S$.  
The generalization of the AGPn case for polarized system is straightforward.

As shown in Ref.~\citenum{Casula:2003p12694}, the evaluation of the wave function $\Psi_{AGP}$, respectively in eqs.~\ref{equ:AGP} or \ref{eq:gagp1}  for the unpolarized or polarized cases,  reduces to the calculation of a single determinant. As a consequence of this, the computational cost for a JAGP evaluation is comparable to that of a JSD calculation.

\subsection{JAGP and JAGPn as a multiconfigurational wave functions}\label{sec:multidetAGP}

In order to simplify the notation, in the following we will refer to  the unpolarized system; the generalization to GAGP is immediate.
We have already seen that the expansion of the pairing function $\cal G$ in terms of MOs  is convenient because it allows one to include chemically meaningful constraints on the wave function that reduce the number of variational parameters, yielding to the JAGPn.
We show here that this expansion also underlines the relation between the AGP and the standard CI-like expansion of the wave function in multiconfigurational approaches.

By substitution of Eq.~(\ref{equ:G_MOs}) in Eq.~(\ref{equ:AGP}), and expanding the summation out of the antisymmetrization operator, the following multi-determinant expansion is obtained for the AGP function:
\begin{eqnarray}\label{equ:multiAGP}
\Psi_{AGP} &=& c_{0} \left| \Psi_0 \right> 
	+ \sum_{i=1}^{N_p} \sum_{a=N_p+1}^{L} c_{ii}^{aa} \left| \Psi_{ii}^{aa} \right> \nonumber \\
	&& +  \mathop{ \sum_{i,j=1 }^{N_p} }_{i\neq j}   \mathop{\sum_{a,b=N_p+1}^{L}}_{a \neq b}
       c_{iijj}^{aabb} \left| \Psi_{iijj}^{aabb} \right> + \ldots 
\end{eqnarray}
where the coefficients are given by:
\begin{equation}\label{equ:coef}
c_0 = \prod_i^{N_p} \lambda_i \,; \qquad 
c_{ii}^{aa} = c_0 {\lambda_a \over \lambda_i} \,; \qquad
c_{iijj}^{aabb} = c_0 {\lambda_a \lambda_b \over \lambda_i \lambda_j } \, ;
\end{equation}
and so on, and 
$\left| \Psi_0 \right> $ is the leading closed-shell Slater determinant:
$$
\left| \Psi_0 \right> = \hat {\cal A} \left\{ 
   \left[ \prod_i^{N_p} \psi_i({\bf r}_i)\alpha(i) \right]
   \left[ \prod_j^{N_p} \psi_j({\bf r}_{N_p+j})\beta(j) \right]
   \right\},
$$
the determinant $\left| \Psi_{ii}^{aa} \right>$ is equal to $\left| \Psi_0 \right> $, but with the virtual orbital $\psi_a$ substituting the valence orbital $\psi_i$, etc.
%
% leading terms 
From the expression of the coefficients in Eq.~(\ref{equ:coef}) and the ordering of the eigenvalues $\lambda_k$ in Eq~(\ref{equ:lambdas}), it follows that the leading contribution beyond the determinant  
$\left| \Psi_0 \right>$
is given by the determinant 
$\left| \Psi_{ii}^{aa} \right>$
with $i=N_p$ and  $a=N_p+1$.
% Confronto con eccitazione singole, doppie, ..., dei metodi di QC: nessuna eccitazione singola, sottoinsieme delle doppie, ... 
The multideterminant expansion of $\Psi_{AGP}$ in Eq.~(\ref{equ:multiAGP}) allows us to directly compare the $\Psi_{AGP}$ with wave functions from other quantum chemical frameworks.
In  $\Psi_{AGP}$ all the odd excited determinants (single, triple, etc.) are excluded, whereas a subset of the even excitations (those with  a multiple excitation to  the same virtual orbital) are taken into account; only doubly occupied molecular orbitals are present.
In other words, $\Psi_{AGP}$ is contained in the seniority zero sector of the electronic full configuration interaction, and its expansion coefficients are determined by the ratios of the eigenvalues of the $\Lambda$ matrix.

The seniority number $\Omega$ represents an alternative tool to classify singlet wave functions. $\Omega$ is defined as the number of unpaired electrons in the Slater determinant, e. g. the number of singly occupied molecular orbitals. $\Omega$-based selection of important Slater determinants in the CI expansion has been seen to be superior than the traditional one, based on the number of excitations with respect to the reference configuration, when the static correlation plays a major role. \cite{Bytautas:2011eo} 
%AGP wave function is characterized by $\Omega=0$, only doubly occupied molecular orbitals are present in Eq. \ref{equ:multiAGP}; 
CI wave functions with $\Omega=0$ for benchmark systems are accurate enough to recover the most part of the static correlation, but the FCI limit (including dynamic correlation) is achieved only when configurations from $\Omega=2, 4, 6...$  sectors are explicitly included.\cite{Bytautas:2011eo}  In the case of JAGP wave function, the combination between a $\Omega=0$ determinantal term and a Jastrow factor allows us to estimate the correlation energy more accurately than  $\Omega=0$ CI wave functions. The set of MOs $\Psi_{k}$ is optimized within the JAGP framework, e. g. in presence of the Jastrow factor and of the multiconfigurational character of the wave function: MOs extracted from our optimization procedure represent therefore the optimal choice for the correlated description of the system under study.

%\subsection{The JAGPn ansatz} \label{sec:JAGPn}

%
The AGPn function can be  expanded in a similar way. 
By substitution of Eq.~(\ref{equ:gn}) in Eq.~(\ref{equ:AGP}) we obtain:
\begin{eqnarray}\label{equ:multiAGPn}
\Psi_{AGPn} &=& c_{0} \left| \Psi_0 \right> 
	+ \sum_{i=1}^{N_p} \sum_{a=N_p+1}^{n} c_{ii}^{aa} \left| \Psi_{ii}^{aa} \right> \nonumber \\
	&&+   \mathop{ \sum_{i,j=1}^{N_p}}_{ i \neq j}   \mathop{  \sum_{a,b=N_p+1}^{n} }_{ a \neq b} 
       c_{iijj}^{aabb} \left| \Psi_{iijj}^{aabb} \right> + \ldots 
\end{eqnarray}
that is different from Eq.~(\ref{equ:multiAGP}) in the fact that the indices $a,b,\ldots$ run from $N_p$ to $n$ (and not to $L$ as for the AGP). 
%
% JSD
In the particular case of $n=N_p$, the  AGPn expansion reduces to the ground state determinant $\left| \Psi_0 \right>$, thus obtaining a closed shell single determinant (SD) wave function.
%When multiplied by a Jastrow factor, the JSD ansatz is defined.
%\footnote{
It is worth stressing out here that the determinantal part of this JSD ansatz corresponds to a restricted calculation, since the up and down-electrons are described by the same MOs. 
Thus, from the point of view of the static correlation, the JSD description is  comparable with approaches like the restricted Hartree-Fock (RHF) or the restricted Kohn-Sham (RKS) DFT.
%}
%
%JAGPn*
It is clear from Eq.~(\ref{equ:multiAGPn}) that the JSD ansatz can be improved by  including in the pairing function ${\cal G}_n$ a number $n>N_p$  of MOs. 
Since typically  JSD  provides an accurate description of  atoms, %(unless the presence of degenerate states gives a considerable contribution, like in the beryllium atom), 
a natural criterium for the choice of the number $n^{*}$ of MOs is by requiring that, when the atoms are at large distances, we cannot obtain an energy below the sum of the JSD atomic energies (defining the JAGPn* wave function, according to \citet{Marchi:2009p12614}).

\subsection{Ionic and Covalent terms in JAGP for the description of Diradicals and Zwitterions }\label{sec:JAGP_diradicals}

We show here that the AGP framework is suitable for the description of the electronic structure of diradical species. Looking at the reduced model, involving two electrons in two molecular orbitals, presented in Ref. \citenum{Salem:1972p92} one obtains three singlet and three triplet wave functions\cite{Salem:1972p92}:
as summarized in Tab.~\ref{tab:salem}, all the three triplet states, and one singlet state, have a leading covalent character, and are thus termed {\em diradical} wave functions, whereas the remaining two singlet states have a leading ionic character and are termed  {\em zwitterionic} wave functions.
%The relative values of the Coulomb repulsion integral $J_{AB}$ between the orbitals, the self repulsion integral $J_{AA}$, and the exchange integral $K_{AB}$ determine the most stable state in a given molecular system, being the energy:
%$J_{AB}+K_{AB}$ for the diradical singlet;
%$J_{AB}-K_{AB}$ for the diradical triplets;
%$(J_{AA}+J_{BB})/2-K_{AB}$ for the zwitterionic 1 singlet;
%$(J_{AA}+J_{BB})/2+K_{AB}$ for the zwitterionic 2 singlet in Tab.~\ref{tab:salem}. 

%\input{tab_salem}
\begin{table*}[htbp]
\caption{ Wave Functions for Diradicals and Zwitterions, according to Salem and Rowland\cite{Salem:1972p92}. 
In this notation, the two odd electrons are localized on the orbitals $\phi_A$ and $\phi_B$, and their overlap is indicated with 
$S_{AB}=\left<\phi_A|\phi_B\right>$.
If the odd orbitals are related by some symmetry element ({\em homosymmetric case})  the proper molecular orbitals are 
$\psi_+ = \frac{\phi_A + \phi_B}{\sqrt{2+2 S_{AB}}} $ and
$\psi_- = \frac{\phi_A - \phi_B}{\sqrt{2-2 S_{AB}}} $. 
If the odd orbitals belong to different symmetry representations of the molecular point group ({\em heterosymmetric case}) the molecular orbitals are $\psi_A$ and $\psi_B$ (though they are not necessarily localized).
The variational parameter $\gamma$ satisfies $1/\sqrt{2} \le \gamma \le 1$.} \label{tab:salem}
{\footnotesize
\begin{tabular}{ l  c  c c }

 & {\bf valence bond approach } & \multicolumn{2}{c}{\bf molecular orbital approach } \\
 &    & {\em homosymmetric case } & {\em heterosymmetric case} \\
\\
\hline
{\bf triplet}$^a$: \\
{\bf \em diradical}     & $ { \phi_A(\textbf {r}_{1}) \phi_B(\textbf {r}_{2}) - \phi_B(\textbf {r}_{1}) \phi_A(\textbf {r}_{2}) } \over \sqrt{2 - 2 {S_{AB}^2}} $ & 
                        $ { \psi_+(\textbf {r}_{1}) \psi_-(\textbf {r}_{2}) - \psi_-(\textbf {r}_{1}) \psi_+(\textbf {r}_{2}) } \over \sqrt{2}$ &  
                        $ { \psi_A(\textbf {r}_{1}) \psi_B(\textbf {r}_{2}) - \psi_B(\textbf {r}_{1}) \psi_A(\textbf {r}_{2}) } \over \sqrt{2}$  \\
\\
\hline
{\bf singlet}$^b$: \\
{\bf \em diradical}      & $ { \phi_A(\textbf {r}_{1}) \phi_B(\textbf {r}_{2}) + \phi_B(\textbf {r}_{1}) \phi_A(\textbf {r}_{2}) } \over \sqrt{2 + 2 {S_{AB}^2}} $ & 
                        $ { \gamma \psi_+(\textbf {r}_{1}) \psi_+(\textbf {r}_{2}) - \sqrt{1-\gamma^2} \psi_-(\textbf {r}_{1}) \psi_-(\textbf {r}_{2}) } \over \sqrt{2}$ &   
                        $ { \psi_A(\textbf {r}_{1}) \psi_B(\textbf {r}_{2}) + \psi_B(\textbf {r}_{1}) \psi_A(\textbf {r}_{2}) } \over \sqrt{2}$   \\
\\
{\bf \em zwitterion 1}   & $ { \phi_A(\textbf {r}_{1}) \phi_A(\textbf {r}_{2}) - \phi_B(\textbf {r}_{1}) \phi_B(\textbf {r}_{2}) } \over \sqrt{2 - 2 {S_{AB}^2}} $  & 
                        $ { \psi_+(\textbf {r}_{1}) \psi_-(\textbf {r}_{2}) + \psi_-(\textbf {r}_{1}) \psi_+(\textbf {r}_{2}) } \over \sqrt{2}$ &  
                        $ { \gamma \psi_A(\textbf {r}_{1}) \psi_A(\textbf {r}_{2}) - \sqrt{1-\gamma^2} \psi_B(\textbf {r}_{1}) \psi_B(\textbf {r}_{2}) } \over \sqrt{2}$ \\
\\
{\bf \em zwitterion 2}   & $ { \phi_A(\textbf {r}_{1}) \phi_A(\textbf {r}_{2}) + \phi_B(\textbf {r}_{1}) \phi_B(\textbf {r}_{2}) } \over \sqrt{2 + 2 {S_{AB}^2}} $ & 
                        $ { \sqrt{1-\gamma^2} \psi_+(\textbf {r}_{1}) \psi_+(\textbf {r}_{2}) + \gamma  \psi_-(\textbf {r}_{1}) \psi_-(\textbf {r}_{2}) } \over \sqrt{2}$ & 
                        $ { \sqrt{1-\gamma^2} \psi_A(\textbf {r}_{1}) \psi_A(\textbf {r}_{2}) + \gamma  \psi_B(\textbf {r}_{1}) \psi_B(\textbf {r}_{2}) } \over \sqrt{2}$  \\
\\
\hline
\multicolumn{4}{ l }{$^a$ the spin part is: $\alpha(1)\alpha(2)$, ${\alpha(1)\beta(2)+\beta(1)\alpha(2)}\over\sqrt{2}$, or $\beta(1)\beta(2)$. } \\
\multicolumn{4}{ l }{$^b$ the spin part is ${\alpha(1)\beta(2)-\beta(1)\alpha(2)}\over\sqrt{2}$. }\\
\end{tabular}
}
\end{table*}

 Starting from such result, a simple model of two electrons and two atomic orbitals $\phi_{A}$ and $\phi_{B}$ centered on nuclei $A$ and $B$ of a chemical system can be  considered a representative scheme for molecules undergoing a bond breaking (like the torsion of C$_{2}$H$_{4}$, the involved atomic orbitals are of $p$ type) or for non bonding electrons (like in CH$_{2}$, where both are centered on the same atom) and, more generally, for those systems characterized by two unpaired electrons. Focusing the attention to the singlet (spin unpolarized) states, the pairing function $\cal G$ term is explicitly written as:
\begin{eqnarray}
{\cal G}  \left( \textbf{r}_{1},\textbf{r}_{2} \right) =& g_{AA} \phi_{A}(\textbf{r}_{1}) \phi_{A}(\textbf{r}_{2}) + g_{BB} \phi_{B}(\textbf{r}_{1}) \phi_{B} (\textbf{r}_{2}) \nonumber \\
 & + g_{AB} \phi_{A}(\textbf{r}_{1}) \phi_{B}(\textbf{r}_{2}) +   g_{BA} \phi_{B}(\textbf{r}_{1}) \phi_{A}(\textbf{r}_{2}) \nonumber \\
\label{eq:two}
\end{eqnarray}
where $g_{\mu \nu}$ coefficients represent the coupling terms of the {\bf G} matrix in the expansion of the AGP spatial factor ( $L=2$ in  Eq.~\ref{equ:calG}): 
\begin{equation}
 \textbf{G} =
\begin{pmatrix}
g_{AA} & g_{AB} \\
g_{BA} & g_{BB}
\end{pmatrix}.
\end{equation}

The elements
$g_{AA}$ and $g_{BB}$ are referred to the ionic terms $\phi_{A}( \textbf{r}_{1}) \phi_{A}( \textbf{r}_{2})$ and $\phi_{B}( \textbf{r}_{1}) \phi_{B} ( \textbf{r}_{2})$ in which the two electrons are localized on the same atom, whereas the elements $g_{AB}$ and $g_{BA}$ (where $g_{AB} = g_{BA}$ for symmetry reasons)  are related to the covalent terms $ \phi_{A}( \textbf{r}_{1}) \phi_{B}( \textbf{r}_{2})$ and $ \phi_{B}( \textbf{r}_{1}) \phi_{A}( \textbf{r}_{2})$.

Following Salem and Rowland's analysis on homosymmetric diradicals, we can rewrite the three singlet states in Tab.~\ref{tab:salem} in terms of $\phi_{A}$ and $\phi_{B}$ atomic orbitals, assuming that $\psi_{+}$ and $\psi_{-}$ molecular orbitals are linear combinations of $\phi_{A}$ and $\phi_{B}$
\begin{eqnarray}
\psi_{+} & = &   \frac  {\phi_{A} + \phi_{B}} { \sqrt{2 + 2\text{S}_{AB}} } \\ 
\psi_{-} & = &   \frac  {\phi_{A} - \phi_{B}} {\sqrt{2 - 2\text{S}_{AB}}  }, \nonumber
\end{eqnarray}
where $\phi_{A}$ and $\phi_{B}$ are related by some symmetry operation.
% ETHYLENE CASE
In the orthogonally twisted ethylene,  S$_{AB} = 0$ and $\gamma=1/\sqrt{2}$, therefore the spatial parts of the wave functions (which have to be multiplied by the singlet spin part ${\alpha(1) \beta(2) - \beta(1) \alpha(2) \over \sqrt{2}}$) become:
\begin{eqnarray}\label{eq:dir1}
\Psi_\textrm{diradical}^\textrm{singlet} (\textbf{r}_{1},\textbf{r}_{2})& = & \frac { \phi_{A}(\textbf{r}_{1})\phi_{B}(\textbf{r}_{2}) +  \phi_{B}(\textbf{r}_{1})\phi_{A}(\textbf{r}_{2}) }{\sqrt{2}}  \\  \nonumber
\Psi_\textrm{zwitterion 1}^\textrm{singlet} (\textbf{r}_{1},\textbf{r}_{2}) & = & \frac { \phi_{A}(\textbf{r}_{1})\phi_{A}(\textbf{r}_{2}) - \phi_{B}(\textbf{r}_{1})\phi_{B}(\textbf{r}_{2}) }{\sqrt{2}}  \\ \nonumber
\Psi_\textrm{zwitterion 2}^\textrm{singlet} (\textbf{r}_{1},\textbf{r}_{2})& = & \frac { \phi_{A}(\textbf{r}_{1})\phi_{A}(\textbf{r}_{2}) + \phi_{B}(\textbf{r}_{1})\phi_{B}(\textbf{r}_{2}) }{\sqrt{2}}  \nonumber
\end{eqnarray}
It is thus evident that 
the singlet ground state $\Psi_\textrm{diradical}^\textrm{singlet}$ is purely covalent, 
whereas $\Psi_\textrm{zwitterion 1}^\textrm{singlet}$ and $\Psi_\textrm{zwitterion 2}^\textrm{singlet}$ are instead ionic states. 

% METHYLENE CASE
In the heterosymmetric case the two molecular orbitals cannot be represented by a linear combination of atomic orbitals, leading to an inversion of the electronic character of the wave functions, as can be seen in Tab.~\ref{tab:salem}. In the methylene CH$_{2}$  the singlet ground state is given by the ionic function: 
\begin{eqnarray}\label{eq:wfCH2}
\Psi_\textrm{zwitterion 1}^\textrm{singlet}(\textbf{r}_{1},\textbf{r}_{2}) & = & 
    \gamma \psi_A(\textbf {r}_{1}) \psi_A(\textbf {r}_{2}) - \nonumber \\
   && \sqrt{1-\gamma^2} \psi_B(\textbf {r}_{1}) \psi_B(\textbf {r}_{2})   ,  
\end{eqnarray}
which  is functionally similar to the purely covalent (diradical) case in homosymmetric systems.  %On the other hand, triplet wave functions assume covalent character. \\

The expansion in Eq. \ref{eq:two} shows that the AGP ansatz contains all the terms reported by the picture in terms of delocalized molecular orbitals and localized atomic orbitals. 
The $g_{\mu \nu}$ coefficients are  variational parameters optimized by the stochastic methods mentioned before and for such reason the AGP optimization is the mandatory step needed to select the right wave function for the ground state of interest.

%%%%%%%%%%%%%%%%%%%%%%%%%%%%%%%%%%%%%%%%%%%%%%%%%%%%%%%%%%%%%%%%%%%%%%
\section{Computational details}
%%%%%%%%%%%%%%%%%%%%%%%%%%%%%%%%%%%%%%%%%%%%%%%%%%%%%%%%%%%%%%%%%%%%%
\label{cdet}

The QMC calculations reported in this paper have been obtained using the {\em TurboRVB} package developed by S. Sorella and coworkers\cite{TurboRVB}, that includes a complete suite of variational and diffusion Monte Carlo codes for wave function and geometry optimization of molecules and solids.
The scalar-relativistic energy consistent pseudopotential ({\bf ECP}) of Burkatzki {\it et al.}\cite{Burkatzki:2007p25447} %Filippi
has been adopted in order to describe the two core electrons of the carbon atoms, whereas the hydrogens are described without  pseudopotential (the nuclear cusp is satisfied by the Jastrow factor, so there is no advantage in using a pseudopotential for the hydrogen).
For the basis sets we have used 
hybrid contracted orbitals\cite{Zen:2013is} constituted by Gaussian type orbitals (GTOs) or mixed GTOs and Slater type orbitals (STOs). The details of the considered  basis sets are reported in Tab.~\ref{tab:basis}. 
The wave function optimization schemes used are the same already described in Ref.\citenum{Zen:2013is}, and all the parameters have been optimized, including the exponents of the basis sets.
% GEO opt
The geometry optimization has been obtained through a 
steepest descent approach, following a method already used successfully for several other molecular systems.\cite{Barborini:2012iy,Barborini:2012it,Coccia:2012kz,Coccia:2012ex}
% diffusion
All the reported  LRDMC results correspond to the continuous extrapolation (lattice mesh size $a\to 0$),  corresponding to the best variational results within the fixed node constraint given by the indicated guiding function.

\begin{table*}[htb]
\caption{ Basis sets used in the QMC calculations reported in the paper, for the determinantal part and the inhomogeneous part of the Jastrow factor.
The number in the curly bracket parenthesis represents the number of contracted hybrid orbitals, as defined in ref. \cite{Zen:2013is}.
}\label{tab:basis}
\begin{tabular}{ l  c c  c c }
Label   & \multicolumn{2}{c}{Determinant $^a$} & \multicolumn{2}{c}{Jastrow $^b$}  \\    
\hline
 {\bf A} & C:(10s,9p,2d,1f)/\{4\} & H:(6s,5p,1d)/\{1\} & C:(4s,2p,1d)/\{2\} & H:(3s,2p)/\{2\} \\ % eA
 {\bf B} & C:(10s,9p,2d,1f)/\{8\} & H:(6s,5p,1d)/\{1\} & C:(4s,2p,1d)/\{2\} & H:(3s,2p)/\{2\} \\ % mA
% {\bf B} & C:(10s,9p,2d,1f)/\{8\} & H:(6s,5p,1d)/\{1\} & C:(4s,2p,1d)/\{2\} & H:(3s,2p)/\{2\} \\ % eB
 {\bf C} & C:(11s,10p,3d,2f)/\{8\} & H:(7s,6p,2d)/\{1\} & C:(4s,2p,1d)/\{2\} & H:(3s,2p)/\{2\} \\ % eC
 {\bf D} & C:(11s,10p,3d,2f)/\{8\} & H:(7s,6p,2d)/\{2\} & C:(4s,2p,1d)/\{4\} & H:(3s,2p)/\{2\} \\ % mB
% {\bf Dc} & C:(11s,10p,3d,2f)/[4s,3p,2d,1f] & H:(7s,6p,2d)/[3s,2p,1d] & C:(4s,2p,1d)/\{4\} & H:(3s,2p)/\{2\} \\ % mB
\hline
\multicolumn{5}{p{15cm}}{ $^a$ GTOs for basis sets {\bf A} and {\bf B}; %{\bf mA}, {\bf eA} and {\bf eB};  
GTOs plus one STO for basis sets {\bf C} and {\bf D}. % {\bf mB} and {\bf eC} . 
} \\
\multicolumn{5}{p{15cm}}{ $^b$ Used uncontracted basis for 3-body Jastrow; hybrid contraction for 4-body Jastrow. } \\
\end{tabular}
\end{table*}

The singlet ground state potential energy surface of C$_{2}$H$_{4}$ has been calculated at VMC and CASSCF level, based on CAS(4,4)//cc-pVDZ structures with a constraint on the torsional angle. For the single point CAS(12,12) energies the cc-pVTZ basis set has been employed. The ORCA package has been used for CASSCF calculations. \cite{orca} \\

%%%%%%%%%%%%%%%%%%%%%%%%%%%%%%%%%%%%%%%%%%%%%%%%%%%%%%%%%%%%%%%%%%%%%
\section{Results and Discussion}
%%%%%%%%%%%%%%%%%%%%%%%%%%%%%%%%%%%%%%%%%%%%%%%%%%%%%%%%%%%%%%%%%%%%%
\label{res}

\subsection{Twisted ethylene}

 Tab~\ref{tab:c2h4comp} collects a selection of our QMC results for the torsional barrier for C$_{2}$H$_{4}$ (presented in more detail below in Tab.~\ref{tab1}) together with  CASSCF calculations and some representative theoretical data available in literature: 
broken symmetry density functional theory\cite{neese2004} (BS-DFT),
spin-flip density functional theory (SF-DFT), 
coupled cluster (CC) methods\cite{krylov98,krylov00,krylov01,Krylov:2006p83,Shao:2003p4807}, 
multireference configuration interaction\cite{Gemein:1996jr,Barbatti:2004p11614} (MRCI), 
and natural orbital functional theory\cite{Lopez:2011p1673} (NOFT). All the latter quantum chemistry methods 
provide estimations of the barrier height within an energy range of 62-73~kcal/mol. From the experimental point of view, Douglas {\it et al.},  studying the kinetics of the thermal cis-trans isomerization of C$_{2}$H$_{4}$ in the temperature range 450-550~\textcelsius, report a value of 65~kcal/mol for the torsional barrier,  \cite{doug55} whereas fitting from resonance Raman spectra of ethylene predicts a value of 60~kcal/mol. \cite{wal89}  \\
Some of the VMC and LRDMC calculations have been carried out on VMC structures optimized using a JAGP/ECP ansatz, see details in Tab.~\ref{tab.geo.c2h4}. 
 The best results, in term of variational energy and variance,  have been obtained by using the basis set {\bf C} in Tab.~\ref{tab:basis}, and provides a torsional barrier of  71.9(1) kcal/mol at VMC level,  and 70.2(2) kcal/mol for LRDMC level. Both estimates are  in very good agreement with results from multi-configurational approaches, like CAS(12,12) (using CAS(4,4)//cc-pVDZ geometries) and MR-CISD+Q, \cite{Barbatti:2004p11614}. Using the JSD wave function, represented by a single closed-shell Slater determinant, similar results to the RHF approach have been obtained, as expected by the lack of the multiconfigurational character.

\begin{table}
\caption{ Ethylene torsional barrier $\Delta$E, computed using CASSCF, VMC, LRDMC  and other representative theoretical approaches: restricted Hartree-Fock (RHF), unrestricted Hartree-Fock (UHF), CCSD, restricted Kohn-Sham DFT (RKS-DFT),  broken-symmetry DFT (BS-DFT), spin-flip DFT (SF-DFT), spin-restricted ensemble-referenced Kohn-Sham DFT (REKS-DFT), natural orbital functional theory  with Piris natural orbital functional (NOFT/PNOF4), and multi-reference configuration-interaction with single and double plus quadruple corrections (MR-CISD+Q).  % from literature.
VMC and LRDMC values have been computed on the VMC/JAGP/ECP structures, and the CAS(12,12) result has been obtained on the CAS(4,4) geometry; see Tab.~\ref{tab.geo.c2h4}. For the other approaches, see the corresponding references. }
\begin{tabular}{l c c}
Approach &  & $\Delta$E [kcal/mol] \\
\hline
RHF   & Ref. \citenum{Lopez:2011p1673} & 108.4 \\
VMC/JSD & This work & 99.3(2) \\
LRDMC/JSD & This work & 97.5(2) \\
CCSD  & Ref. \citenum{Casanova:2008ef} & 89.9 \\
UHF & Ref. \citenum{krylov00} & 49.6 \\
RKS-DFT/BLYP & Ref. \citenum{Filatov:1999bu} & 88.7 \\
REKS-DFT/BLYP & Ref. \citenum{Filatov:1999bu} & 69.1 \\
SF-DFT/B3LYP & Ref. \citenum{Shao:2003p4807} & 79.6 \\
%BS-DFT/M06-2X & Ref. \citenum{Lopez:2011p1673} & 66.0 \\
BS-DFT/B3LYP & Ref. \citenum{Lopez:2011p1673} & 63.2 \\
NOFT/PNOF4 & Ref. \citenum{Lopez:2011p1673} & 73.2 \\
CAS(12,12)  & This work & 69.1 \\
VMC/JAGP & This work & 71.9(1) \\
LRDMC/JAGP & This work & 70.2(2) \\
MR-CISD+Q & Ref. \citenum{Barbatti:2004p11614} &  69.2 \\
%\hline
%Expt.$^{a}$  & Ref. \citenum{doug55} & 64.6 \\
%Expt. $^{b}$ & Ref. \citenum{wal89} & 59.7 \\
%
%\multicolumn{3}{l}{ $^a$ Kinetics of the {\it cis-trans} isomerization at 450-550 \textcelsius }  \\
%\multicolumn{3}{l}{ $^b$ Fitting from resonance Raman spectra } 
\end{tabular}
\label{tab:c2h4comp}
\end{table}

Tab.~\ref{tab1} shows that 
the choice of the basis set in the VMC calculations has a very small effect on the convergence of $\Delta E$, with a difference of  $\simeq$1 kcal/mol when moving from the basis set {\bf A} to {\bf C} (see the results for the CAS(4,4) geometries). Even with the smallest basis set, {\bf A}, with only four hybrid orbitals on the carbon atoms, a good estimate of the torsional energy has been obtained. Furthermore, differences in the VMC $\Delta E$ due to the employed geometry are negligible.
On the other hand, Tab.~\ref{tab1} underlines the failure of the single determinant JSD wave function in the proper description of the torsional barrier profile, similarly with what found by other single-reference methods with restricted orbitals, as RHF and  CCSD calculations\cite{Casanova:2008ef}, whereas unrestricted HF calculations are seen to underestimate the barrier. 
The static electron correlation problem is challenging also for DFT approaches, indeed restricted Kohn-Sham DFT calculations are seen to 
overestimate the torsional barrier and predict a wrong  sharp cusp for a torsion of 90 degrees\cite{Filatov:1999bu} (similarly to single-reference methods with restricted orbitals),
while spin-flip DFT and broken-symmetry DFT approaches produce a better agreement with multireference findings. 
Dynamical correlation, introduced by the presence of the Jastrow factor, plays a minor role in the estimation of the barrier height, as expected;  JSD strongly overestimates the barrier height, with VMC values of 102.9(1) and 99.3(2) kcal/mol, when using the  basis set {\bf C}. The first result has been obtained by projecting the optimized JAGP function into the JSD ansatz, whereas in the second case the JSD wave function parameters have been re-optimized. The application of LRDMC (97.5(2) kcal/mol) does not alter the above picture indicating that the nodal surface coming from the optimization of a single determinant wave function is not correct. Specifically, the JSD total energy, at VMC and LRDMC level, is about 50 mHartrees above the JAGP energies for the orthogonally twisted structure; while, for the planar geometry the total energy is underestimated by about 10 (VMC) or 5 (LRDMC) mHartrees with respect to the JAGP values. This indicates, once again, that a single-reference approach, even in the QMC framework, cannot be successfully used for systems characterized by a strong static correlation as at the top of the torsional barrier. \\
The LRDMC correction in the barrier height (see Tabs.~\ref{tab:c2h4comp} and \ref{tab1}) of less than 2 kcal/mol also confirms the good quality of the fully optimized trial wave function $\Psi_{T}$. This encouraging result makes us confident that the variational flexibility and the protocols employed in the optimization of the variational parameters can lead to a correlated high-level wave function $\Psi_{T}$ and, consequently, to a reliable VMC description of the electronic structure.

\begin{table*}
\caption{ Energies (in Hartrees) of the planar and orthogonally twisted ethylene, for JSD and JAGP wave functions, computed using VMC and LRDMC($a\to 0$), and the corresponding torsional barrier $\Delta E$ (in kcal/mol). 
The geometrical parameters, CAS and JAGP, are reported in Tab.~\ref{tab.geo.c2h4}. 
The basis are described in Tab.~\ref{tab:basis}.%), and are obtained using the ECP pseudopotential\cite{Burkatzki:2007p25447} for the C atoms.   
}
\begin{tabular}{ l c c  c c  c }
ansatz	&geo.	&	basis	&	planar	&	twisted at 90\degree	& $\Delta E$ \\ 
		&		&			& [Hartrees] & [Hartrees] & [kcal/mol] \\
\hline
                                                                                                                                                                          
VMC/JSD $^a$	 & JAGP & {\bf C} &	-13.7199(2)  &	-13.5560(2)  &	102.9(1) \\ % JSD(proj)/VTZ+STO-HYB-8-1-Jnew-2-2/GEOopt	

VMC/JSD		 & JAGP & {\bf C} &	-13.7198(2)  &	-13.5616(2)  &	99.3(2) \\ % JSD(opt)/VTZ+STO-HYB-8-1-Jnew-2-2/GEOopt

VMC/JAGP		 & CAS  & {\bf A} &	-13.7260(2) 	 &	-13.6103(2)  &	72.7(2) \\ % HYB_4_1_Jnew_2_2 OPTvmc_8
VMC/JAGP		 & CAS  & {\bf B} &	-13.7289(1)  &	-13.6144(2)  &	71.8(1) \\ % HYB_8_1_Jnew_2_2 OPTvmc_9
VMC/JAGP		 & CAS  & {\bf C} &	-13.7292(1)  &	-13.6150(2)  &	71.6(1)  \\ % +STO_HYB_8_1_Jnew_2_2 OPTvmc_9

VMC/JAGP		 & JAGP & {\bf C} &	-13.7299(2)  &	-13.6154(2)  &	71.9(1) \\ % /VTZ+STO-HYB-8-1-Jnew-2-2/GEOopt

\hline

%LRDMC(0.5)/JSD(opt)/VTZ+STO-HYB-8-1-Jnew-2-2/GEOopt	&	-13.7477068531702       1.4354284E-04		-13.5910590990331       1.5501957E-04		98.2969   0.132573
%LRDMC(0.3)/JSD(opt)/VTZ+STO-HYB-8-1-Jnew-2-2/GEOopt	&	-13.7452303833695       1.3770349E-04		-13.5895249722548       1.4623866E-04		97.7056   0.126045
%LRDMC(0.2)/JSD(opt)/VTZ+STO-HYB-8-1-Jnew-2-2/GEOopt	&	-13.7443821477807       1.4099189E-04		-13.5891231110148       1.5202088E-04		97.4255   0.130105
%LRDMC(0.1)/JSD(opt)/VTZ+STO-HYB-8-1-Jnew-2-2/GEOopt	&	-13.7439197015790       2.6654408E-04		-13.5881721221804       3.0991787E-04		97.7321   0.256506
%LRDMC(fit)/JSD(opt)/VTZ+STO-HYB-8-1-Jnew-2-2/GEOopt	&	-13.7437193159837       2.432602767163595E-004	-13.5883903667162       2.719584308770348E-004	97.4694   0.228963
LRDMC/JSD	& JAGP & {\bf C} &	-13.7437(2)  &	-13.5884(3)  &	97.5(2) \\ % fit a->0  JSD(opt)/VTZ+STO-HYB-8-1-Jnew-2-2/GEOopt

%VMC/JAGP/VTZ-HYB-4-1-Jnew-2-2/GEOcas44			&	-13.7260(2)  &	-13.6103(2)  &	72.7(2) \\
%VMC/JAGP/VTZ-HYB-8-1-Jnew-2-2/GEOcas44			&	-13.7289(1)  &	-13.6144(2)  &	71.8(1) \\
%VMC/JAGP/VTZ+STO-HYB-8-1-Jnew-2-2/GEOcas44		&	-13.7292(1)  &	-13.6150(2)  &	71.6(1) \\

%LRDMC(0.5)/JAGP/VTZ+STO-HYB-8-1-Jnew-2-2/GEOopt	&	-13.7501(1)  &	-13.6383(1)  &	70.1(1) \\
%LRDMC(0.3)/JAGP/VTZ+STO-HYB-8-1-Jnew-2-2/GEOopt	&	-13.7490(1)  &	-13.6374(1)  &	70.0(1) \\
%LRDMC(0.2)/JAGP/VTZ+STO-HYB-8-1-Jnew-2-2/GEOopt	&	-13.7488(1)  &	-13.6369(1)  &	70.2(1) \\
%LRDMC(0.1)/JAGP/VTZ+STO-HYB-8-1-Jnew-2-2/GEOopt	&	-13.7483(2)  &	-13.6366(2)  &	70.1(2) \\
LRDMC/JAGP	& JAGP & {\bf C} &	-13.7484(2)  &	-13.6364(2)  &	70.2(2) \\ % fit a->0   /VTZ+STO-HYB-8-1-Jnew-2-2/GEOopt

\hline
\multicolumn{6}{l}{ $^a$ JSD function obtained from the projection of the optimized JAGP function. }

\end{tabular}
\label{tab1}
\end{table*}

\begin{figure}[htbp]
\caption{
{\bf Upper panel}: 
Ethylene torsional barrier, in kcal/mol, for CASSCF(12,12)/cc-pVTZ (black solid circles) and VMC/JAGP/ECP (red squares) on the structures obtained by CASSCF(4,4)/cc-pVDZ. 
The barriers height for the LRDMC$_{a\to 0}$/JAGP/ECP (green diamond), VMC/JSD/ECP (blue up triangle) and LRDMC$_{a\to 0}$/JSD (violet down triangle) are also reported.
All the QMC results are obtained using the C basis, see Tab.~\ref{tab:basis}.
{\bf Lower panel}:
Plot of the ratio (red squares)
${ (c_{66}^{77} / c_0) = {\lambda_{7}}/{\lambda_{6}} }$
between the coefficients of the leading two determinants in the JAGP wave function -- see Eq.~(\ref{equ:multiAGP}) -- as a function of the torsional angle.
All other determinants in the JAGP have a relative weight 
$|\lambda_a/\lambda_i| < 0.03$.
The ratio is zero for the JSD wave function (blue line), by definition.
For a comparison, also the corresponding ratio of the coefficients for the CAS(12,12) calculation is reported.
}\label{fig_C2H4_S}
\includegraphics[width=0.47\textwidth]{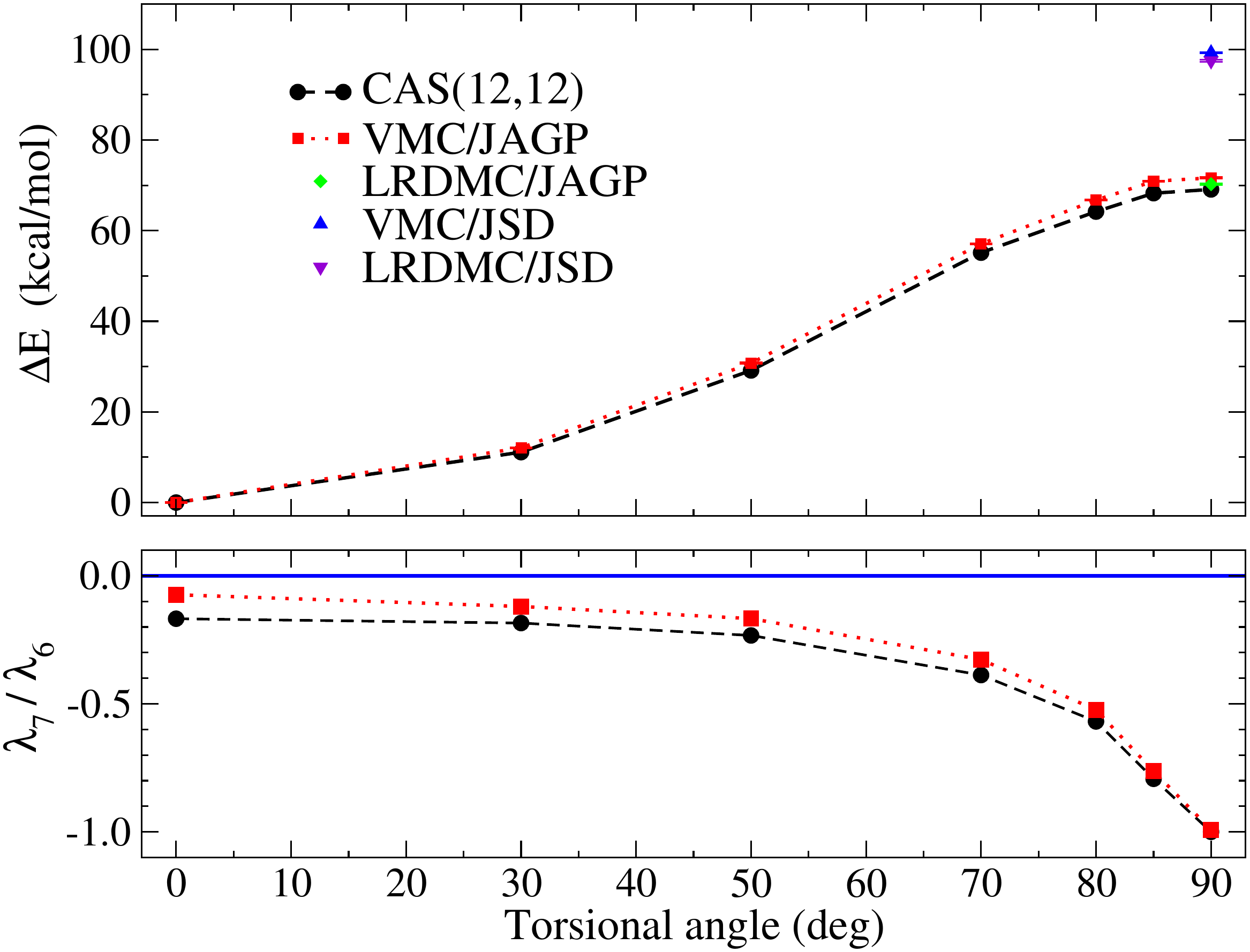}
\end{figure}

\begin{figure}[htbp]
\caption{ Molecular orbitals $\psi_6$ and $\psi_7$  -- corresponding to  $\alpha=N_p$ and $N_p+1$ in Eq.~(\ref{equ:G_MOs}) -- of the VMC/JAGP/ECP calculation of the planar and twisted configurations of the singlet ethylene.  }
\label{fig_C2H4_MOs}
\includegraphics[width=0.45\textwidth]{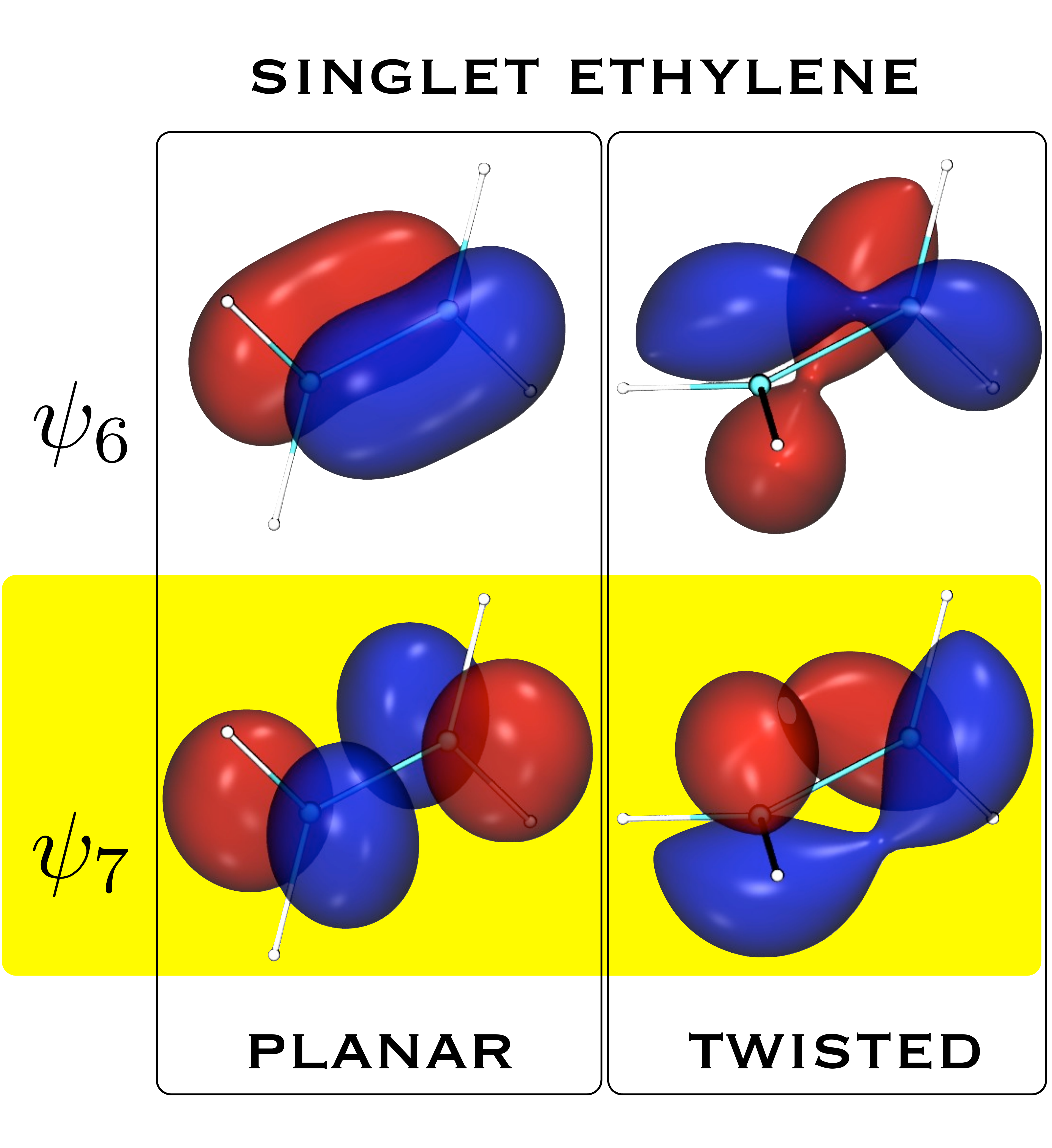}
\end{figure}

A deeper insight in the description of the multiconfigurational nature of the JAGP wave function has been made possible thanks to the analysis of the relative weight of the determinants in the expansion of the $\Psi_{AGP}$ (Eq. \ref{equ:multiAGP}) along the torsional energy profile (see Fig.~\ref{fig_C2H4_S}).   In the orthogonally twisted configuration the two frontier $\pi$ and $\pi^{*}$ orbitals become degenerate and the two electronic configurations, $(\pi)^{2}$ and $(\pi^{*})^{2}$, assume the same weight. In the simplified model \cite{Salem:1972p92}, the singlet ground state wave function is given by the term $ [\gamma \psi_{+}^{2} - \sqrt{1-\gamma^{2} } \psi_{-}^{2}  ]$ in Tab.~\ref{tab:salem}, with $\gamma=1/\sqrt{2}$ and $\psi_{\pm}^{2} = \psi_{\pm}(\textbf {r}_{1}) \psi_{\pm}(\textbf {r}_{2})$ ($\psi_{+} = \pi$ and $\psi_{-} = \pi^{*}$ in the case of the twisted C$_{2}$H$_{4}$). \\ 
The lower panel of Fig.~\ref{fig_C2H4_S} reports the behaviour of the ratio between $\lambda_{7}$ and $\lambda_{6}$, eigenvalues from the generalized eigenvalue problem in Eq. \ref{gendiag}, as a function of the dihedral angle. These two parameters are related to the $\psi_{6,7}$ molecular orbitals (see Eq. \ref{equ:G_MOs}) that are strictly related to the two frontier orbitals in the traditional picture. Following the definitions in  Eq.~\ref{equ:coef} for the coefficients of the expansion of $\Psi_{AGP}$ into single electron determinants, it can be easily verified that  $\lambda_{7}/\lambda_{6}$ corresponds to the ratio between the coefficient of the ground state determinant $\Psi_{0}$ and that of the doubly excited determinant $\Psi_{66}^{77}$, where two electrons move to the first ``virtual'' orbital $\psi_{7}$. The absolute value of such ratio converges to -1 when increasing the angle, clear evidence that the $\Psi_{AGP}$ wave function is dominated by two configurations with the same weight for the singlet ground state of the orthogonally twisted ethylene. In other words, the multiconfigurational nature of the wave function is fully recovered by $\Psi_{AGP}$, in a similar fashion as what observed in traditional multi determinant approaches. Furthermore, the coefficients corresponding to the other double excitations are much smaller and do not give an appreciable contribution to the $\Psi_{AGP}$ expansion.  The minus sign in the ratio directly derives from the definition of the wave function (see Tab.~\ref{tab:salem}). 
The fully optimised JAGP $\psi_{6}$ and $\psi_{7}$ molecular orbitals with {\bf C} basis set are displayed in Fig.~\ref{fig_C2H4_MOs}. As expected, for the twisted conformation, the $\pi$ and $\pi^{*}$ orbitals are identical but rotated.

%\begin{table}
%\caption{Energy differences in kcal/mol, for the torsion around the C=C bond in C$_2$H$_4$, computed on the CAS(4,4)/cc-pVTZ structures  using CASSCF(12,12)/cc-pVTZ, VMC/JAGP and VMC/JSD.
%XXX QUESTA TAB. LA METTEREI SUL SUPP.INFO., TANTO C'E' GIA' LA FIGURA XXX
%}
%  \begin{tabular}{c|c|c|c}
%  & CASSCF(12,12) & VMC/JAGP & VMC/JSD   \\
%\hline
%30\degree & 11.10 & 12.1(1) &  \\
%50\degree & 29.15 & 30.8(1) &  \\
%70\degree & 55.12 & 57.1(1) &  \\
%80\degree & 64.20 & 66.7(1) &  \\
%85\degree & 68.26 & 70.9(1) &  \\
%90\degree & 69.09 & 71.6(1) & 99.3(2) \\
%\end{tabular}
%\label{tab1}
%\end{table}

\begin{table}
\caption{Geometrical parameters calculated for the singlet ground state ethylene %C$_{2}$H$_{4}$
for CASSCF(4,4)/cc-pVTZ and for VMC/JAGP with the basis set {\bf C} in Tab.~\ref{tab:basis}, and ECP pseudopotential\cite{Burkatzki:2007p25447}.  
Bond lengths are reported  in \AA~ and angles in deg.}
\begin{tabular}{c|c|c}
  & CAS & JAGP \\
\hline
planar  &  &  \\
C=C   & 1.340  & 1.3293(8) \\
%C-H  & 1.083  &  \\
H-C-H & 116.86 & 116.90(7) \\
C-C-H & 121.57 & 121.55(4) \\
\hline
twisted at 90\degree~  &  &  \\
C-C   &  1.468  & 1.4514(8) \\
%C-H  &  1.083  & \\
H-C-H & 116.86  & 116.80(8) \\
C-C-H & 121.57  & 121.60(4) \\
\hline
%\multicolumn{3}{l}{ $^a$ basis: C XXX Jas. XXX, H XXX Jas. XXX }
\end{tabular}
\label{tab.geo.c2h4}
\end{table}

Moving from planarity to the orthogonally twisted ethylene obviously  results in an increase of the carbon-carbon length (see Tab.~\ref{tab.geo.c2h4}): breaking of the $\pi$ bond induces a stretching well described from both CAS(4,4)//cc-pVTZ  (0.128 \AA) and VMC/JAGP/ECP with basis set {\bf C}   (0.122(1) \AA) calculations. Lack of dynamic correlation in the CASSCF wave function is responsible for the slightly larger bond distances. Bond angles do not vary moving from the planar to the orthogonally twisted structure and do not essentially depend on the methodology used.

Comparison between VMC/JAGP/ECP singlet and triplet barrier height profiles shows the expected cross of the curves in the proximity of 90\textdegree, the triplet state becoming more stable with a singlet-triplet gap of about 2 kcal/mol: variational energies have been computed on the singlet CAS(4,4)//cc-pVDZ structures from 0\textdegree~ to 90\textdegree. Such finding, which follows the theoretical statement reported in Ref. \citenum{Salem:1972p92}, according to which the singlet-triplet gap is proportional to the exchange interaction, is in agreement with the fact that the triplet state should be more stable in energy than the singlet one for such diradical system; MRCI calculations\cite{Gemein:1996jr} indeed predict the triplet state lower in energy at 90\textdegree, representing a further confirmation of the reliability of the JAGP ansatz not only in the description of the multiconfigurational character of the singlet state (in the twisted conformation), but also of the triplet state, as written in Eq. \ref{eq:gagp1}.

\begin{figure}[htbp]
\caption{
VMC/JAGP singlet and triplet torsion barriers, calculated on singlet CAS(4,4)//cc-pVDZ structures, using ECP pseudopotential and the basis set {\bf C} in Tab.~\ref{tab:basis}. 
}\label{fig:TS-C2H4}
\includegraphics[width=0.5\textwidth]{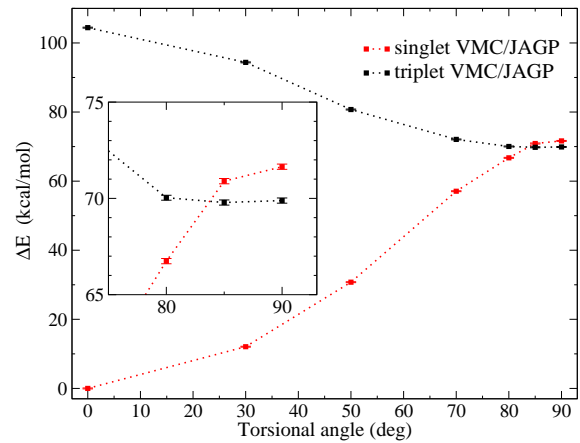}
\end{figure}

\subsection{Methylene}

As mentioned in the introduction, 
methylene is  an example of a heterosymmetric diradical,
according to Salem and Rowland\cite{Salem:1972p92} (see Tab.~\ref{tab:salem}).

%%%%INTRO%%%
The ground state of CH$_{2}$ is the triplet ($\tilde{X} ^{3}B_{1}$), but the lowest singlet state ($\tilde{a} ^{1}A_{1}$) is quite close in energy, as it is typical for diradical species. The reference experimental measurement for $T_{0}$ obtained by fitting rotation-vibration data gives a value of 8.998 kcal/mol (0.390 eV). \cite{jensen88} Furthermore,  the relativistic and nonadiabatic effects calculated in Refs.~\citenum{dav80,handy86}, and described in Ref.~\citenum{piecuch94} have been taken into account and added to the adiabatic gap $T_{e}$, 9.215 (0.400 eV): the final result is 9.363 kcal/mol (0.406 eV). 
%Indeed, the experimental estimate is of 9.363 kcal/mol (0.406 eV) \cite{dav80,handy86,jensen88}, corrected for non-adiabatic (Born-Oppenheimer diagonal correction) and relativistic effects, as described in Refs \citenum{dav80,handy86}.

In Tab.~\ref{tab:ch2_compare} we report a list of the results obtained for the adiabatic singlet-triplet energy gap of the methylene by other quantum chemical approaches or experimentally derived, in comparison with the most representative QMC results obtained in this work. 
Highly accurate approaches yield
theoretical estimates that compare well with the experimentally derived values, giving an energy range, depending on the specific method applied,  from 8.56 kcal/mol (0.371~eV) for CMRCI+Q and complete basis set (CBS) extrapolation\cite{woon95}, to 11.25 kcal/mol (0.488 eV) for CCSDT//TZ2P \cite{shen12}. \\Excellent agreement is also found by \citet{Zimmerman:2009hh} using a Quantum Monte Carlo (QMC) approach on the FCI structures.\cite{Sherrill:1998p1040} Starting from wave functions derived from CAS(2,2), CAS(4,4) and CAS(6,6) active spaces, that account for the static correlation of the system, a Jastrow factor is introduced, in order to also include the dynamic correlation. Accurate results are obtained both in the variational and fixed node diffusion Monte Carlo schemes\cite{Zimmerman:2009hh} and by Anderson and Goddard \cite{and2010}, who performed diffusion Monte Carlo calculations using a Generalized Valence Bond (GVB) wave function. \\
%
%%expt. (jensen88)-> 0.390 o 0.406 eV   rotation-vibration data and fitting  \\ 
%%          -> 8.998 kcal/mol (T0, 0.390 eV) e 9.215 kcal/mol (Te, 0.400 eV) \\
%%alijah90 ->  \\          
%%piecuch94 ->   CCSD 0.411 (Te) \\
%%woon95 ->     RCCSD(T)/CBS 0.411, CMRCI/CBS 0.399, CMRCI+Q 0.371 (AE) 	\\	
%%Yamaguchi:1996p7911 -> CISD//TZ3P(2f,2d) 12.19 and CASSCF SOCI//TZ3P(2f,2d)  9.47 kcal/mol \\
%%Sherrill:1998p1040 -> 0.483 eV (FCI) \\
%%Slipchenko:2002p4694 -> spin-flip approaches
%%shen08 ->  0.398 (CAS-BCCC4/cc-pVQZ), 0.421 (MR-CISD/cc-pVQZ), 0.451 (SF-OD/cc-pVQZ) \\
%%Zimmerman:2009hh -> 0.407(8)-0.430(8) (VMC), 0.406(4)-0.412(4) (DMC) \\
%%gour10 ->           0.467 (EOM-CCSD//CBS), 0.430 (CR-EOMC(2,3)//CBS) \\
%%shen12 ->  0.488 (CCSDT//TZ2P),  0.413 (CCSDT//aug-cc-pVTZ), 0.401 (CC(t;3)$_{D}$//CBS)  \\    
The effect of pseudopotential (ECP) on the singlet-triplet gap of CH$_{2}$ diradical results in a slight underestimation  \cite{woon95,Zimmerman:2009hh} of about 0.02 eV (0.46 kcal/mol) \cite{woon95} at RCCSD(T)//CBS and MRCI and 0.03 eV (0.69 kcal/mol) at VMC and DMC level \cite{Zimmerman:2009hh}. The shift from Ref. \cite{Zimmerman:2009hh} has been calculated by averaging the gap on the different wave functions adopted. Pseudopotential in QMC calculations by Zimmerman {\it et al.} \cite{Zimmerman:2009hh} is the same used by us in the present work;  the  small red shift observed in our results with respect to the experimental measure or to the best theoretical estimates could be due to the pseudopotential, and this fact make us confident of the convergence and accuracy of our calculations. \\
%It is worth to point out the effect of pseudopotential on the estimation of the adiabatic ($\tilde{X} ^{3}B_{1}$ - $\tilde{a} ^{1}A_{1}$) gap: using CBS extrapolation, the gap is underestimated of about 0.5 kcal/mol (0.02 eV)  at RCCSD(T), CMRCI and CMRCI+Q level; \cite{woon95} the red shift is about 0.7 kcal/mol (0.03 eV), on average for the different wave functions adopted, for QMC, with respect to the corresponding all-electron calculations. \cite{Zimmerman:2009hh} 
%(same pseudopotential used in the present work) and in \cite{woon95}:
%%%woon 95 ->  RCCSD(T)/CBS 0.392, CMRCI/CBS 0.379, CMRCI+Q 0.389 (valence electrons)  ->  singlet-triplet underestimated of 0.02 eV \\
%%\citet{Zimmerman:2009hh} -> 0.361(8) - 0.396(8) (VMC), 0.362(2) - 0.388(4) (DMC) -> red-shift of almost 0.04 eV \\
%pair functions \cite{Ellis:2013p2857} \\

\begin{table}
\caption{ Comparative table of the 
adiabatic gap $\Delta E = E( \tilde{X} ^{3}B_{1} ) - E( \tilde{a} ^{1}A_{1} )$ 
for CH$_2$, calculated using different approaches on the FCI structures from Ref.~\cite{Sherrill:1998p1040}. }
\label{tab:ch2_compare}
\begin{tabular}{l c c}
Approach &  & $\Delta$E [kcal/mol] \\
\hline

VMC/JSD/ECP   & this work & 	13.45(10)  \\ % VMC/JSDopt/filVTZ+STO,C8J4,H2J2/GeoLET
%VMC/J$^*$SD/ECP$^b$ & this work &	13.73(9)   \\ % path: CH2/VTZ+STO_h8,2_J4,2_XXX/JSD_optDFT/OPT_Jas_spincontaminated
LRDMC/JSD/ECP    & this work & 	13.36(8)  \\ % LRDMC(fit)/JSDopt/filVTZ+STO,C8J4,H2J2/GeoLET

CASSCF(6,6) & this work & 10.53 \\
%VMC/J$\cdot$CAS(2,2)/ECP & Ref.~\cite{Zimmerman:2009hh} & 8.9(2) \\ % 0.386(8)
%DMC/J$\cdot$CAS(2,2)/ECP & Ref.~\cite{Zimmerman:2009hh} & 8.88(9) \\ % 0.385(4)
%VMC/J$\cdot$CAS(4,4)/ECP & Ref.~\cite{Zimmerman:2009hh} & 8.3(2) \\ % 0.361(8)
%DMC/J$\cdot$CAS(4,4)/ECP & Ref.~\cite{Zimmerman:2009hh} & 8.35(9) \\ % 0.362(4)
DMC/GVB  & Ref.~\citenum{and2010} & 9.4(1) \\
VMC/J$\cdot$CAS(6,6) & Ref.~\citenum{Zimmerman:2009hh} & 9.9(2) \\ % 0.430(8)
DMC/J$\cdot$CAS(6,6) & Ref.~\citenum{Zimmerman:2009hh} & 9.36(9) \\ % 0.406(4)

VMC/J$\cdot$CAS(6,6)/ECP & Ref.~\citenum{Zimmerman:2009hh} & 9.1(2) \\ % 0.396(8)
DMC/J$\cdot$CAS(6,6)/ECP & Ref.~\citenum{Zimmerman:2009hh} & 8.95(9) \\ % 0.388(4)

VMC/JAGP/ECP & this work &  8.09(8) \\ % VMC/filVTZ+STO,C8J4,H2J2/GeoLETT
LRDMC/JAGP/ECP & this work & 	8.58(7) \\ % LRDMC(fit)/filVTZ+STO,C8J4,H2J2/GeoLETT

VMC/J$^*$AGP/ECP$^b$ & this work &  8.32(7) \\ % VMC/filVTZ+STO,C8J4,H2J2/GeoLETT
LRDMC/J$^*$AGP/ECP$^b$ & this work & 	8.64(6) \\ % LRDMC(fit)/filVTZ+STO,C8J4,H2J2/GeoLETT

RCCSD(T) & Ref.~\citenum{Woon:1995ee} & 9.48 \\ % 0.411
%SU/SS MRCCSD & Ref.~\cite{} & 9.48 \\ % 0.411  P. Piecuch, X. Li, and J. Paldus, Chem. Phys. Lett. 230, 377 1994.
CMRCI & Ref.~\cite{Woon:1995ee} & 9.18 \\ % 0.398
CMRCI+Q & Ref.~\citenum{Woon:1995ee} & 8.97 \\ % 0.389
FCI & Ref.~\cite{Sherrill:1998eg} & 11.12 \\ % 0.482
\hline
Expt. T$_{0}$ & Ref.~\citenum{jensen88} & 8.998 \\
Expt. T$_{e}$ & Ref.~\citenum{jensen88} & 9.215 \\
Expt.$^{a}$ T$_{e}$ & Ref.~\citenum{jensen88} & 9.363 \\
\hline
\multicolumn{3}{l}{ $^a$ Relativistic and nonadiabatic corrections from Refs.~\citenum{dav80,handy86} } \\
\multicolumn{3}{p{8.5cm}}{ $^b$  2-body Jastrow  satisfying the cusp condition for pairs of electrons both of like and  unlike spin  (details in Section~\ref{sec:Jas} and in Tab.~\ref{tab:ch2_qmc}). } \\ %iesdrr=-7 Two body 1/2b(1?e?br) one body rescaled, + cusp for parallel spins.
\end{tabular}
\end{table}

Methylene is an atypical  diradical, because its lowest energy singlet  has  {\em zwitterion 1} wave function, in the Salem and Rowland's two electrons model reported in Tab.~\ref{tab:salem}.
This  unusual behavior is mainly due to the fact that the unpaired electrons are centered on the same carbon atom.
%%%%INTRO%%%%
 We therefore need an ansatz  describing both the triplet {\em diradical}  and the singlet {\em zwitterion 1}  functions (Tab.~\ref{tab:salem}) in order to accurately estimate the adiabatic energy gap. For the former case a single Slater determinant is enough to properly describe the electronic structure, while for the latter a multiconfigurational wave function must be used.

% table results qmc
%\input{Tab_ch2}
\begin{table*}
\caption{ Evaluation of the adiabatic gap 
$\Delta E = E(\tilde{X} ^{3}B_{1}) - E(\tilde{a} ^{1}A_{1})$  
at VMC, DMC or LRDMC level, using ECP pseudopotential\cite{Burkatzki:2007p25447} for the C atom. Two geometries are considered, one from a FCI approach (Ref.~\citenum{Sherrill:1998p1040}), and one from VMC/JAGP/ECP approach (this work, see Tab.~\ref{tab:ch2_geo}). The basis sets are defined in Tab.~\ref{tab:basis}. % (A) and filVTZ+STO,C8J4,H2J2 (B). 
Three different ansatzes are here compared: JSD, JAGP and J$\cdot$CAS(6,6) with ECP; the former two calculated in this work, the latter taken from Ref.~\citenum{Zimmerman:2009hh}. J$^*$ indicates the use of the 2-body Jastrow defined in Eq.\ref{equ:2BJas-sc},  satisfying  the cusp condition both for like and  unlike-spin pairs of electrons. 
}\label{tab:ch2_qmc}
\begin{tabular}{ l c c  l l  l l }

Method   & geo.	& basis		&	$\tilde{X} ^{3}B_{1}$				&	$\tilde{a} ^{1}A_{1}$				&	$\Delta$E [eV]	&	$\Delta$E [kcal/mol] \\

\hline

%VMC/JSD $^a$ & FCI & {\bf D} &	-6.7152(1)  &	-6.6966(1)  &	0.508(4) & 	11.71(10)  \\ % VMC/JSDproj/filVTZ+STO,C8J4,H2J2/GeoLET

VMC/JSD      & FCI & {\bf D} &	-6.7195(1)  &	-6.6981(1)  &	0.583(4) & 	13.45(10)  \\ % VMC/JSDopt/filVTZ+STO,C8J4,H2J2/GeoLET

VMC/J$^*$SD $^a$ & FCI & {\bf D} &	-6.7227(1)  &	-6.7008(1)  &	0.595(4) &	13.73(9)   \\ % path: CH2/VTZ+STO_h8,2_J4,2_XXX/JSD_optDFT/OPT_Jas_spincontaminated

%\hline

VMC/JAGP		 & FCI & {\bf A} &	 -6.7220(1)	&	-6.7086(1)	&		0.365(4)  &	8.41(9)  \\ 

VMC/JAGP		 & FCI & {\bf B} &	-6.72299(9)  & 	-6.71035(9)  & 		0.344(4)  & 	7.93(8)  \\ % VMC/filVTZ/GeoLETT
VMC/JAGP		& JAGP & {\bf B} &	-6.72306(10) & 	-6.71046(9)  & 		0.343(4)  & 	7.91(8)  \\ % VMC/filVTZ/GeoVMC

VMC/JAGP     & FCI & {\bf D} &	-6.72335(9)  & 	-6.71046(8)  &		0.351(3)  &  8.09(8) \\ % VMC/filVTZ+STO,C8J4,H2J2/GeoLETT

VMC/J$^*$AGP $^a$ & FCI & {\bf D} & 	-6.72525(8)  &	-6.71200(8)  &		0.361(3)  &	8.32(7)  \\ % CH2/VTZ+STO_h8,2_J4,2_XXX/JAGP_Jas_spincontaminated

%VMC/JAGP     & FCI & Dc &	-6.72250(9)  & 	-6.70788(10)  &		0.398(4)  &  9.17(8) \\ % VMC/filVTZ+STO,contratti,symmAGP/GeoLETT

VMC/J$\cdot$CAS(6,6) $^b$ & FCI & Ref.~\citenum{Zimmerman:2009hh} & -6.7251(2) & -6.7105(2) & 0.396(8) & 9.1(2) \\ % Needs

\hline

%LRDMC(0.5)/JSDopt/filVTZ+STO,C8J4,H2J2/GeoLET	-6.73111183859183       9.2912000E-05		-6.70882361733971       9.0000016E-05		0.606493 0.00351992 	13.9859 0.0811705
%LRDMC(0.3)/JSDopt/filVTZ+STO,C8J4,H2J2/GeoLET	-6.72986588836904       8.3336286E-05		-6.70837707834169       8.2495753E-05		0.58474  0.00319087 	13.4843 0.0735825
%LRDMC(0.2)/JSDopt/filVTZ+STO,C8J4,H2J2/GeoLET	-6.72926680953141       8.3904415E-05		-6.70817874394588       8.3481493E-05		0.573835 0.00322074 	13.2328 0.0742713
%LRDMC(0.1)/JSDopt/filVTZ+STO,C8J4,H2J2/GeoLET	-6.72934877813800       8.0695485E-05		-6.70795650839743       8.0612015E-05		0.582113 0.00310378 	13.4237 0.071574
LRDMC/JSD    & FCI & {\bf D} &	-6.72920(9) &	-6.70791(9)  &	0.5793(4) & 	13.36(8)  \\ % LRDMC(fit)/JSDopt/filVTZ+STO,C8J4,H2J2/GeoLET

%LRDMC(0.5)/filVTZ/GeoVMC		-6.73129074048290       7.4994532E-05		-6.71775291033087       7.3037219E-05			0.368383 0.00284858 	8.49503 0.0656892
%LRDMC(0.3)/filVTZ/GeoVMC		-6.73101779080773       7.2689225E-05		-6.71728946822241       7.0141774E-05			0.373567 0.0027487 	8.61456 0.0633859
%LRDMC(0.2)/filVTZ/GeoVMC		-6.73087681696642       7.0789894E-05		-6.71719380044051       6.4046850E-05			0.372334 0.00259768 	8.58613 0.0599034
%LRDMC(0.1)/filVTZ/GeoVMC		-6.73075049276277       7.2194387E-05		-6.71709666543503       6.7703426E-05			0.37154 0.00269321 	8.56782 0.0621063
LRDMC/JAGP		& JAGP & {\bf B} &	-6.73072(8)  & 	-6.71708(8)  & 		0.371(3)  & 	8.55(7)  \\ % LRDMC(fit0)/filVTZ/GeoVMC

%LRDMC(0.5)/filVTZ+STO,C8J4,H2J2/GeoLETT	-6.73121355456382       7.5657430E-05		-6.71765388824004       7.1574083E-05			0.368977 0.00283402 	8.50873 0.0653535
%LRDMC(0.3)/filVTZ+STO,C8J4,H2J2/GeoLETT -6.73106885371066       7.3200572E-05		-6.71730705012884       6.8942456E-05	                0.374478 0.00273625 	8.63557 0.0630988
%LRDMC(0.2)/filVTZ+STO,C8J4,H2J2/GeoLETT	-6.73084152148211       7.0900016E-05		-6.71721762840428       6.8646390E-05			0.370725 0.00268541 	8.54903 0.0619264
%LRDMC(0.1)/filVTZ+STO,C8J4,H2J2/GeoLETT	-6.73083598834168       7.0425049E-05		-6.71710726940662       6.7554392E-05			0.373577 0.00265549 	8.61481 0.0612363
LRDMC/JAGP & FCI & {\bf D} &	-6.73077(8)  &	-6.71709(8)  &		0.372(3)  & 	8.58(7) \\ % LRDMC(fit)/filVTZ+STO,C8J4,H2J2/GeoLETT

LRDMC/J$^*$AGP $^a$ & FCI & {\bf D} & -6.73084(7)	&	-6.71707(6)	&	0.375(3)  &	8.64(6) \\

DMC/J$\cdot$CAS(6,6) $^b$ & FCI & Ref.~\citenum{Zimmerman:2009hh} & -6.7308(1) & -6.7165(1) & 0.388(4) & 8.95(9)  \\ % Needs

\hline
%\multicolumn{7}{p{15cm}}{ $^c$ JSD function obtained from the projection of the optimized JAGP function. } \\
%\multicolumn{7}{p{17cm}}{ $^a$  The total squared spin  $\left< S^2 \right>$ for the singlet and triplet functions is respectively: 0.0049(4) and 2.0038(3)  for VMC/J$^*$SD, 0.0030(3) and 2.0029(3) for VMC/J$^*$AGP, 0.000(1) and 2.000(1) for LRDMC/J$^*$AGP. } \\ %iesdrr=-7 Two body 1/2b(1?e?br) one body rescaled, + cusp for parallel spins.
\multicolumn{7}{p{17cm}}{ $^a$  The total squared spin  $\left< S^2 \right>$ for the singlet and triplet functions are reported in Table~\ref{tab:ch2_s2}. } \\ 
\multicolumn{7}{p{17cm}}{ $^b$ From Ref.~\citenum{Zimmerman:2009hh}. }\\
\end{tabular}
\end{table*}

\begin{table}
\caption{ Value of the total squared spin $\left< S^2 \right>$ for the spin contaminated wave functions considered in Tab.~\ref{tab:ch2_qmc}.
}\label{tab:ch2_s2}
\begin{tabular}{ l c c   }
% & \multicolumn{2}{c}{ $\left< S^2 \right>$ } \\
% & singlet  & triplet \\
&	$\tilde{X} ^{3}B_{1}$				&	$\tilde{a} ^{1}A_{1}$	\\
% &	$\tilde{a} ^{1}A_{1}$  & $\tilde{X} ^{3}B_{1}$ \\
\hline
VMC/J$^*$SD & 2.0038(3) & 0.0049(4)   \\
VMC/J$^*$AGP &2.0029(3)& 0.0030(3)   \\
LRDMC/J$^*$AGP &  2.000(1) & 0.000(1)  \\
\end{tabular}
\end{table}

Details of QMC results shown in Tab.~\ref{tab:ch2_qmc} allow one to get a deeper insight into the performance on CH$_{2}$ of the computational procedures presented in this work:  singlet and triplet energies, the corresponding adiabatic gap, using both the VMC and LRDMC, for the Jastrow correlated single Slater determinant, with or without spin-contamination (see Section~\ref{sec:Jas}) denoted in the table respectively with JSD and J$^*$SD, and the Jastrow correlated antisymmetrized geminal power, indicated with JAGP and J$^*$AGP. 
% commento spin contamination
The total squared spin $\left< S^2 \right>$, 
 of the wave 
 functions J$^*$SD and J$^*$AGP can be efficiently evaluated as described in Appendix~\ref{app:s2}, and are  reported in Tab.~\ref{tab:ch2_s2}.
Two different geometries, and different basis set are compared in the Tab.~\ref{tab:ch2_qmc}. In all our calculations, as for the ethylene, we use the ECP pseudopotential for the two core electrons of the carbon atom.
For a further comparison, in Tab.~\ref{tab:ch2_qmc} we also report the VMC and DMC results obtained by \citet{Zimmerman:2009hh} for a Jastrow correlated CAS(6,6) ansatz.

% commenti su JSD
The most evident conclusion that we extract from Tab.~\ref{tab:ch2_qmc} is that the JSD ansatz (and also J$^*$SD) is unable to accurately describe the singlet methylene, overestimating the adiabatic singlet-triplet energy gap by about 4~kcal/mol, i.e., more than 40\%. 
Also in this case the LRDMC is unable to significantly correct the inaccuracy coming from the single closed-shell Slater determinant wave function.
In particular, we observe that LRDMC energy of the triplet JSD is only $\sim$1.5~mH higher than the more accurate (multideterminant) JAGP and J$\cdot$CAS ansatzes, whereas the singlet JSD is $\sim$9~mH higher than the corresponding  JAGP or  J$\cdot$CAS energies.
%Thus, for the singlet molecule, the poor description of the static correlation in the JSD function is reflected on an inaccurate nodal surface.
%Therefore, the complexity of this system calls for a more accurate ansatz for the determinantal part of the wave function, unless a more sophisticated and computationally demanding released QMC approach is used.

% good JAGP results
The JAGP ansatz seems instead to provide a very accurate description of the electronic structure both of the singlet and of the triplet, and with a comparable computational effort, as discussed in Ref.~\citenum{Zen:2013is}.
Indeed, in Tab.~\ref{tab:ch2_qmc} we observe that the JAGP results are all comparable to the J$\cdot$CAS(6,6) results by \citet{Zimmerman:2009hh}.
In particular, VMC/JAGP calculations with the smallest considered basis set {\bf A} 
are $\sim$3~mH higher than the J$\cdot$CAS(6,6) for the triplet, and $\sim$2~mH for the singlet
whereas the results with basis set {\bf B} and {\bf D} have almost the same energy of J$\cdot$CAS(6,6) for the singlet, and are no more than $\sim$2~mH
higher for the triplet.
We observe for the JAGP results a weak dependence on the basis set size, that is more evident in the triplet than in the singlet.
The use of the hybrid contracted orbitals, introduced in Ref.  \citenum{Zen:2013is}, 
greatly simplifies the convergence of the basis set, and in particular we obtain reliable results having used 8 hybrids for the  carbon atom and 2 for the hydrogen atom, and some STOs in order to get the correct tails (basis set {\bf D}).
The accuracy of the JAGP ansatz is confirmed by the close LRDMC results.

\begin{figure}[htbp]
\caption{Leading valence ($\psi_{3}$ and $\psi_{4}$) and non bonding ($\chi_{1}$ and $\chi_{2}$) molecular orbitals of the VMC/JAGP/ECP of methylene for the singlet and triplet states, see Eq.~\ref{eq:gagp1}.}
\label{fig:ch2_orb}
\includegraphics[width=0.45\textwidth]{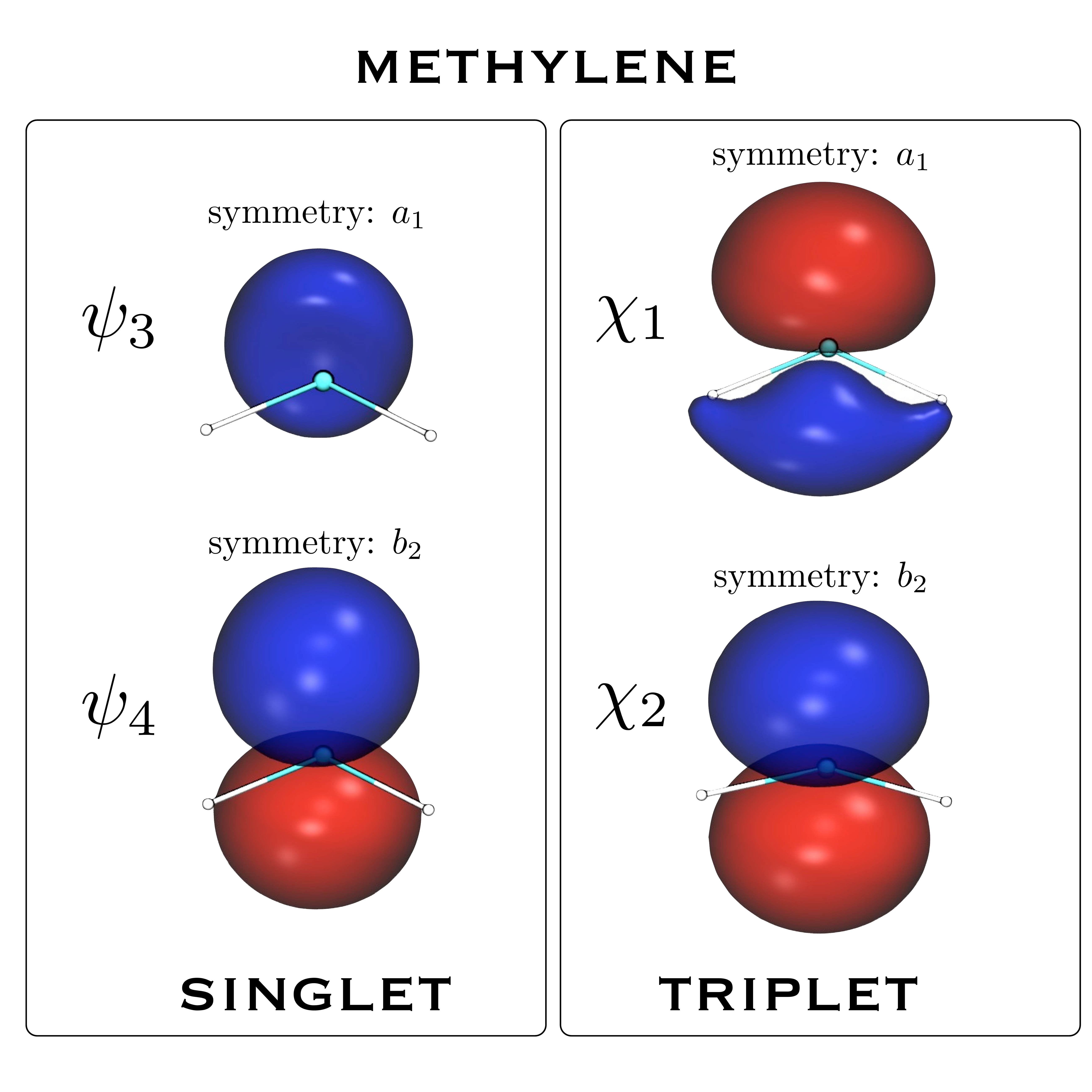} 
\end{figure}

% ansatz of the singlet
In order to better understand how the JAGP includes static correlation in the system, we have to consider the multideterminant expansion of the AGP in Eq.~\ref{equ:multiAGP}.
The question on how many determinants are necessary to reach a given accuracy finds a clear answer, for the singlet state, in Tab.~\ref{tab:ch2_ansatz}
%
%We could ask how many determinants are really necessary to reach a given accuracy.
%This can be answered, for the singlet, considering the Tab.~\ref{tab:ch2_ansatz}
, where the JAGP has been compared to the JSD, JDD (a Jastrow correlated double determinant constructed according to Salem and Rowland's model), and JAGPn with different values of $n$.
The JSD wave function looses more than 10 mH ($\simeq 6.3$ kcal/mol).
The JDD ansatz, instead,  is 4.32(4)~mH ($\simeq 2.7$ kcal/mol) higher than the JAGP (and  the energy difference likely decreases if the parameters of the JDD are variationally optimized).
Thus, in agreement with Salem and Rowland, two leading determinants are dominant in the expansion of Eq. \ref{equ:multiAGP}, constructed using the two quasi-degenerate heterosymmetric orbitals reported in Fig.~\ref{fig:ch2_orb}.
Then, by considering larger values of {\it n}, for the JAGPn, the overlap with the JAGP increases with a consequent decrease in the energy difference. In particular, the JAGPn* introduced in Ref.~\citenum{Marchi:2009p12614} is only  $<$2~mH ($\simeq 1.3$ kcal/mol) higher than the JAGP.

% TAB confronto ansatz per CH2
%\input{Tab_ch2_ansatz.tex}
\begin{table}
\caption{ 
Comparison of different ansatzes for the 
singlet methylene, in the FCI geometry, with the ECP pseudopotential and the basis set {\bf D} of Tab.~\ref{tab:basis}.  
The wave function overlap and the energy difference $\Delta E$, in mH and kcal/mol, with a reference JAGP function $\Psi_\textrm{JAGP}$ with a VMC energy of -6.71046(8) H are reported.
Both $\Delta E$ and the overlap (defined as ${\left< \Psi_\textrm{JAGP} | \Psi_\textrm{Ansatz} \right>}^2$) are computed using the correlated sampling technique.
Unless specified, the considered wave functions have been obtained by projection of the AGP part into the truncated AGPn, without further optimization of the parameters.
According to Section~\ref{sec:AGP}, JSD correspond to JAGPn with n=3, and JAGPn$^*$ to n=6.
}
\begin{tabular}{ l c c c  c c  c }
Ansatz	& &	$\Delta E$ [mH]	& $\Delta E$ [kcal/mol] & ${\left< \Psi_\textrm{JAGP} | \Psi_\textrm{Ansatz} \right>}^2$	 \\
\hline
JAGPn$^a$ 	& n=12 &  $< 10^{-8}$  & $< 6 \times 10^{-9}$ & $\sim 1$ \\
JAGPn 	& n=10 &  0.006(1)  &  0.0038(6)  &0.99999874(5) \\
JAGPn* 	& n=6  &  0.27(1)   &0.169(6) &  0.999884(1) \\
JAGPn 	& n=5  &  1.80(3)   & 1.80(2) &0.99876(1)  \\
JAGPn 	& n=4  &  3.49(4)   &2.19(3) & 0.99763(1)  \\
JDD$^b$	&      &  4.32(4)   & 2.71(3) &0.99661(2)  \\
JSD  	&      &  14.1(1)   & 8.85(6) &0.9716(2)  \\
JSD$^c$ &      &  11.8(4)   & 7.4(3) &0.9740(2)  \\

\hline
\multicolumn{6}{p{8.5cm}}{ $^a$ In the multideterminat expansion of the AGPn, see Eqs.~\ref{equ:multiAGPn} and \ref{equ:coef}:
${\lambda_4 / \lambda_3} \sim -0.152$; 
${\lambda_5 / \lambda_3} \sim -0.036$;
${\lambda_4 / \lambda_2} \sim -0.034$;
${\lambda_4 / \lambda_1} \sim -0.031$;
${\lambda_6 / \lambda_3} \sim -0.030$;
all the others are $|{\lambda_a / \lambda_i} |< 0.01$.
} \\
\multicolumn{6}{p{8.5cm}}{ $^b$ Jastrow correlated Double Determinant, according to Salem and Rowland model, see Tab.~\ref{tab:salem}.  } \\
\multicolumn{6}{p{8.5cm}}{ $^c$ the parameters of the wave function, both of the determinant and of the Jastrow, have been optimized within the JSD ansatz. }
\end{tabular}
\label{tab:ch2_ansatz}
\end{table}

% delta E
The adiabatic singlet-triplet energy gap predicted by the JAGP is between 7.91(8) and 8.64(6)~kcal/mol, depending on the basis set and on the QMC method used. 
%However, there is an evidence that the largest basis, or the use of LRDMC in place of VMC, gives the largest values for $\Delta$E.
 The LRDMC on the largest basis ({\bf D}, with STO functions) results in the best agreement with VMC results from Ref. \citenum{Zimmerman:2009hh}.
% JAGP triplet
Thus, the results seem to indicate that the JAGP is a slightly better ansatz for the singlet than for the triplet wave function.
This actually is  not  surprising, because in the triplet we use the GAGP in Eq.~\ref{eq:gagp1}, with two parallel-spin electrons described by a couple of unpaired functions. But a description of the triplet as accurate as that for the singlet would be obtained by using for the two parallel-spin electrons a pairwise antisymmetric function, analogous to the geminal used to describe the singlet pair (actually, this ansatz is the Pfaffian function\cite{Bajdich:2006p18510,Bajdich:2008p18507}). In other words, the two unpaired electrons should be described by an antisymmetric $L\times L$ coupling matrix, thus by $L(L-1)/2$ independent parameters, and not $2L$ as in the GAGP. 

% J*
Moreover, there is another aspect that differentiates the singlet from the triplet: the number of like-spin and unlike-spin pairs.
Six valence electrons are present in CH$_{2}$. One can easily verify that the triplet state contains seven like-spin pairs instead of six, as in the singlet state.
%On the 15 electrons pairs, in the singlet 9 are unlike-spin and 6 are like-spin; whereas in the triplet 8 are unlike-spin and 7 are like-spin.
As discussed in Section~\ref{sec:Jas}, 
 the Jastrow factor usually treats only the unlike-spin electron-electron cusps, in order to avoid the spin contamination of the wave function.
However the presence of an extra like-spin electron couple in the triplet could bias the energy difference favoring the singlet, as actually observed.
Calculations have been carried out using the homogeneous two-body Jastrow of Eq.~\ref{equ:2BJas-sc}, and the results are reported in Tab.~\ref{tab:ch2_qmc} labeled by J$^*$. 
As expected the triplet energy experiences the largest improvement,  $\sim $2~mH, leading to the increase of the adiabatic energy gap.
Moreover, the total spin squared $\left<S^2\right>$, reported in the Tab.~\ref{tab:ch2_s2}, demonstrate that the level of spin contamination is negligible, and converges to zero in the LRDMC. 
Nevertheless, the J$^*$ slightly improves the results only in the VMC scheme, because in the LRDMC the results are almost identical within the stochastic error.

Geometry optimization by VMC using JAGP, pseudopotential on the carbon atom and the {\bf B} basis set for triplet ($\tilde{X} ^{3}B_{1}$) and first singlet ($\tilde{a} ^{1}A_{1}$) states of CH$_{2}$ shows a very good agreement with FCI//TZ2P calculations: \cite{Sherrill:1998p1040} the main difference is observed for the C-H bond of the singlet state, about 3 m\AA~ (see Tab.~\ref{tab:ch2_geo}).

\begin{table}
\caption{C-H bond length (in \AA) and H-C-H angle (in deg) for triplet ($\tilde{X} ^{3}B_{1}$) and singlet ($\tilde{a} ^{1}A_{1}$) states of CH$_2$ (see text for details) obtained by FCI \cite{Sherrill:1998p1040} and VMC/JAGP/ECP calculations, for the basis set {\bf B} in Tab.~\ref{tab:basis}. }
\begin{tabular}{l c | l l |  l l   }
&& \multicolumn{2}{c}{ $\tilde{X} ^{3}B_{1}$ } & \multicolumn{2}{c}{  $\tilde{a} ^{1}A_{1}$} \\
method & & C-H & H-C-H  & C-H & H-C-H \\ 
\hline
FCI & ref.\cite{Sherrill:1998p1040} & 1.0775 & 133.29 & 1.1089 & 101.89 \\
%\hline
JAGP & this work & 1.0748(1) & 132.89(2) & 1.1062(2) & 101.97(1)  \\
\hline
%\multicolumn{6}{l}{ $^a$ basis: XXX }
\end{tabular}
\label{tab:ch2_geo}
\end{table}

%%%%%%%%%%%%%%%%%%%%%%%%%%%%%%%%%%%%%%%%%%%%%%%%%%%%%%%%%%%%%%%%%%%%%
\section{Conclusions}
%%%%%%%%%%%%%%%%%%%%%%%%%%%%%%%%%%%%%%%%%%%%%%%%%%%%%%%%%%%%%%%%%%%%%
\label{con}

A well-balanced characterization of the static electronic correlation is of primary importance in order to capture the qualitative features in systems like diradicals and transition states.
Multi-configurational quantum chemistry methods are able to provide a reliable description of the electronic correlation in diradical species.
Unfortunately the price to pay for these approaches is given by the computational cost rapidly increasing with the system size, thus limiting the application range. 
As shown by Scuseria and coworkers,  \cite{Bytautas:2011eo} a selection of configurations with small seniority number (i.e., the number of unpaired electrons in a determinant) leads to a fast convergence of the CI expansion in cases where static correlation dominates the electronic structure. 
Through a detailed investigation of the multiconfigurational nature of the AGP  we have shown how this ansatz is a zero seniority function with some constraints on the coefficients of the expansion and with molecular orbitals optimized at fully correlated level. 

%%%%%
In this work we have investigated two archetypal diradical systems using a JAGP ansatz within a QMC approach: the torsion of the ethylene and the $\tilde{X} ^{3}B_{1}$ - $\tilde{a} ^{1}A_{1}$ gap of methylene. The Jastrow factor efficiently takes into account the dynamical correlation of the system. 
The determinant part of the function, the AGP, is responsible for recovering the static correlation.
The results obtained using the JAGP ansatz on C$_{2}$H$_{4}$ and CH$_{2}$ demonstrate that even the simplest and computationally cheapest VMC scheme is sufficient to quantitatively describe diradicals states. Indeed the computationally more expensive fixed-node projection scheme (LRDMC) produces a rigid shift of the absolute energies, but the energy differences are only slightly affected.
On the other hand, the JSD ansatz, i.e. a single determinant correlated with a Jastrow factor, produces  inaccurate results, also using the LRDMC approach. The reason of such failure is coming from  the static correlation that represents the leading ingredient for the electronic structure of diradical species, and a poor description of it dramatically affects the quality of the nodal surface of the wave function. 
%As a matter of fact, reliable results starting from a JSD function could be achieved only using a released node QMC approach.
The moderate scaling with respect to the system size ($N^{d}$, with $3<d<4$ and $N$ the number of electrons) and the availability of High Performance Computing facilities allows one to successfully carry out QMC calculations on larger molecules.\cite{Coccia:2012kz}
This work can be considered a fundamental step for the application of JAGP-based QMC methods in the study of diradical species of chemical and biological interest, and, generally, in systems in which static correlation plays an essential role.

%%%%%%%%%%% Acknowledgments
\section*{Acknowledgement}

%\begin{acknowledgement}
The authors thank Matteo Barborini for valuable discussions. The authors acknowledge funding provided by the European Research Council project n. 240624 within the VII Framework Program of the European Union. Computational resources were supplied by CINECA, PRACE infrastructure, and the Caliban-HPC centre at the University of L'Aquila.
%\end{acknowledgement}

%\input{S^2.tex}

\newcommand\zenx{\bar{ \mathbf{x}}}
\newcommand\zenr{\mathbf{r}}
\newcommand\zensigma{\bar{ \mathbf{\sigma}}}
\newcommand\np{{N_p}}
\newcommand\ssq{{S^2}}
\newcommand\sz{{S_z}}
\newcommand\cg{{\cal G}}
\newcommand\bfa{{\bf A}}

\appendix

\section{Efficient calculation of \boldmath$\langle \ssq \rangle$}\label{app:s2}

In this appendix we describe how to compute the expectation value of the total spin square $\ssq$ over the variational 
wave function $\Psi_{JAGP}$ in the paired AGP case, namely with vanishing spin projection $\sz$ along the $z$-axis and $N=2\np$ electrons.

As is well known, within Variational Monte Carlo, we need to compute the so called local estimator 
of the spin square:
$$ \frac{\langle \zenx | \ssq | \Psi \rangle}{\langle \zenx | \Psi \rangle}$$
where $\zenx = \{ \zenr_1^\uparrow , \ldots , \zenr_\np^\uparrow , \zenr_1^\downarrow , \ldots, \zenr_\np^\downarrow \}$ is a many body configuration where the electron positions and the spin projection along the z-axis $\sigma_i = \pm 1/2 $ 
are defined.
 The application of $\ssq$  to a given configuration can be written as:
\begin{equation}
  \ssq  | \zenx \rangle = \sz^2 | \zenx \rangle +\frac{1}{2}\sum_{\zenr_i,\zenr_j}(S^+_{\zenr_i} S^-_{\zenr_j}+{\rm h.c.}) | \zenx \rangle \label{spin2} 
\end{equation}
where $i$ and $j$ label all electron positions, regardless of their spins.
The above expression can be recast in the following way:
\begin{equation}
 \label{ssq}
  \ssq | \zenx \rangle = - \sum_{k,l}^\np | \zenx_{kl} \rangle  + \frac{N}{2} | \zenx \rangle 
\end{equation}
which generates $\np^2$ new configurations
\begin{equation} \label{xij}
  | \zenx_{kl} \rangle  = - S^-_{\zenr_k^\uparrow} S^+_{\zenr_l^\downarrow} |\zenx \rangle \qquad (k,l=1,\ldots,\np)
\end{equation}
where  $k$($l$) labels only the spin-up(down) electrons, namely the new configuration $\zenx_{kl}$ is obtained by swapping the positions of the (k,l) electron pair 
with opposite spins.
The minus sign in the above expression takes into account the Fermi statistics, in order to recast a spin-flip with a position exchange.
Similarly  the rightmost term in Eq.~(\ref{ssq}) takes into account the local term $i=j$ in Eq.(\ref{spin2}), which is obtained by applying the 
spin-flip operator to each individual electron, leading to a trivial constant ($N/2$) times $|\zenx \rangle$.

Therefore, the main problem is to compute the $\np^2$ wave function ratios:
\begin{eqnarray}
r_{kl}&=&  \frac{\langle \zenx_{kl} | \Psi \rangle }{ \langle \zenx | \Psi \rangle } = \frac{ \det \bfa'}{ \det \bfa} \frac{J'}{J} \\
     \bfa_{ij}&=&\cg(\zenr_i^{\uparrow},\zenr_j^{\downarrow}) 
\end{eqnarray}
which contain a determinant factor and a Jastrow factor.

\subsection{Determinant part}
For the determinant factor, the swapping of the (k,l) electron pair implies a change in the determinant $\bfa \to \bfa' $ given by: 
{\setlength\arraycolsep{2pt}
  \begin{eqnarray}
     \bfa'_{ij}&=&\bfa_{ij}+\delta_{ik}[\cg(\zenr_l^{\downarrow},\zenr_j^{\downarrow})-\cg(\zenr_k^{\uparrow},\zenr_j^{\downarrow})]  \
                   +\delta_{jl}[\cg(\zenr_i^{\uparrow},\zenr_k^{\uparrow})-\cg(\zenr_i^{\uparrow},\zenr_l^{\downarrow})] \nonumber \\
           &&\hphantom{A_{ij}}+\delta_{ik}\delta_{jl} \theta_{kl} \nonumber \\
         \theta_{kl}&=&[\cg(\zenr_l^{\downarrow},\zenr_k^{\uparrow})+\cg(\zenr_k^{\uparrow},\zenr_l^{\downarrow})  \
                    -\cg(\zenr_l^{\downarrow},\zenr_l^{\downarrow})-\cg(\zenr_k^{\uparrow},\zenr_k^{\uparrow})]  \nonumber
  \end{eqnarray}
}

We rewrite $\bfa'$ as
  \begin{eqnarray}
     \bfa'&=&\bfa(I+\Delta)  \nonumber \\
     \Delta_{ij}&=&\bfa^{-1}_{ik}W_j+\delta_{jl}U_i  \nonumber \\
     W_j&=&\cg(\zenr_l^{\downarrow},\zenr_j^{\downarrow})-\cg(\zenr_k^{\uparrow},\zenr_j^{\downarrow})  \nonumber \\
     U_i&=& B^{\uparrow,\uparrow}_{i,k} -B^{\uparrow,\downarrow}_{i,l} + \bfa^{-1}_{ik} \theta_{kl}\nonumber 
%     U_i&=&\sum_z \bfa^{-1}_{iz}[\cg(r_{z\uparrow},\zenr_k^{\uparrow})-\cg(r_{z\uparrow},\zenr_l^{\downarrow})+\delta_{zk}\otimes]  \nonumber 
  \end{eqnarray}
where we have defined the  $B^{\uparrow,\sigma}$ matrices as follows:
\begin{equation}
B^{\uparrow,\sigma}_{i,j}= \sum_z \bfa^{-1}_{iz}\cg(\zenr_z^\uparrow,\zenr_j^\sigma) 
\end{equation}
Notice that these matrices can be computed only once for all spin flip ratios, amounting to $2N_p^3$ operations. 

Then, by employing the Sherman-Morrison algebra, the ratio of the two determinants  $\det \bfa^\prime/ \det \bfa $  is given by  a determinant of 
a much simpler  $2 \times 2$ matrix $M$:
\begin{eqnarray}
   {\bf M}&=&\left( 
   \begin{array}{cc}
     1+ \sum_i \bfa^{-1}_{ik}W_i & \bfa^{-1}_{lk} \\
   \sum_i  U_i W_i   & 1+U_l 
   \end{array}
   \right)
\end{eqnarray}

In this way, the number of operations necessary to obtain all the $\np^2$ ratios scales as $N^3$ namely, with a computational 
time similar to the calculation of the energy.

\subsection{Jastrow part}

If the wave function is defined in terms of a spin dependent  two-body jastrow, the total Jastrow factor  can be generally  written as
\begin{equation}
J = \exp \bigg( \sum_{i<j}^N V(i,j) \bigg)
\end{equation}
where the summation over $i$ and $j$ are now over all the electrons, regardless of their spins, and $V$ is defined as
\begin{eqnarray} \label{jas2b}
  V(i,j)&=&\frac{1}{2}[V_{ee}(i,j)(1+4\sigma_i \sigma_j)+2 V_{ee}(i,j)(1-4\sigma_i \sigma_j)] \nonumber \\
        &=&\frac{1}{2}V_{ee}(i,j)(3-4\sigma_i \sigma_j)
\end{eqnarray}
where $V_{ee}$ is defined as
\[ V_{ee}(i,j)= \frac{ 1-\exp({- b r_{ij}}) }{ 4 b } \]
which is consistent with Eq.~(\ref{equ:2BJas-sc}). The spin dependent part of $V$ is rewritten as $V_{sd} (i,j) \sigma_i \sigma_j$, $  V_{sd} (i,j) = -2V_{ee}(i,j)$.
Each time we swap the electron $k$ with spin up and the electron $l$ with spin down, we only need to flip the corresponding spins $\sigma_k$ and $\sigma_l$ 
in Eq.~(\ref{jas2b}).
It is clear therefore that, by computing 
the auxiliary vector ${\bar V}(l)=\sum_{i\ne l} V_{sd}(i,l)\sigma_i$ once for all, all the Jastrow ratios $J'/J$ can be easily computed as
\[ \frac{J'}{J}=\exp\{[{\bar V}(l) - V_{sd}(k,l)/2] - [{\bar V}(k) + V_{sd}(l,k)/2] \}\,, \]
namely also this with an irrelevant number of operations ( $ \simeq N^2$ operations).

%%%%%%%%%%% Bibliography
%\bibliography{Bibliography}

\end{document}